# Bimodal Distribution of Sulfuric Acid Aerosols in the Upper Haze of Venus

Peter Gao[1,]*, Xi Zhang[1,2], David Crisp[3], Charles G. Bardeen[4], and Yuk L. Yung[1]

[1]*Division of Geological and Planetary Sciences, California Institute of Technology, Pasadena, CA, USA, 91125*

[2]*Department of Planetary Sciences and Lunar and Planetary Laboratory, University of Arizona, Tucson, AZ, USA, 85721*

[3]*Jet Propulsion Laboratory, California Institute of Technology, Pasadena, CA, USA, 91109*

[4]*National Center for Atmospheric Research, Boulder, CO, USA, 80307*

*Corresponding author at: *Division of Geological and Planetary Sciences, California Institute of Technology, Pasadena, CA, USA, 91125*

*Email address:* pgao@caltech.edu

*Phone number:* 626-298-9098

**Abstract**

Observations by the SPICAV/SOIR instruments aboard Venus Express have revealed that the upper haze (UH) of Venus, between 70 and 90 km, is variable on the order of days and that it is populated by two particle modes. We use a one-dimensional microphysics and vertical transport model based on the Community Aerosol and Radiation Model for Atmospheres to evaluate whether interaction of upwelled cloud particles and sulfuric acid particles nucleated *in situ* on meteoric dust are able to generate the two observed modes, and whether their observed variability are due in part to the action of vertical transient winds at the cloud tops. Nucleation of photochemically produced sulfuric acid onto polysulfur condensation nuclei generates mode 1 cloud droplets, which then diffuse upwards into the UH. Droplets generated in the UH from nucleation of sulfuric acid onto meteoric dust coagulate with the upwelled cloud particles and therefore cannot reproduce the observed bimodal size distribution. By comparison, the mass transport enabled by transient winds at the cloud tops, possibly caused by sustained subsolar cloud top convection, are able to generate a bimodal size distribution in a time scale consistent with Venus Express observations. Below the altitude where the cloud particles are generated, sedimentation and vigorous convection causes the formation of large mode 2 and mode 3 particles in the middle and lower clouds. Evaporation of the particles below the clouds causes a local sulfuric acid vapor maximum that results in upwelling of sulfuric acid back into the clouds. In the case where the polysulfur condensation nuclei are small and their production rate is high, coagulation of small droplets onto larger droplets in the middle cloud may set up an oscillation in the size modes of the particles such that precipitation of sulfuric acid "rain" may be possible immediately below the clouds once every few Earth months. Reduction of the polysulfur condensation nuclei production rate destroys this oscillation and reduces the mode 1 particle abundance in the middle cloud by two orders of magnitude. However, it better reproduces the sulfur-to-sulfuric-acid mass ratio in the cloud and haze droplets as constrained by fits to UV reflectivity data. In general we find satisfactory agreement between our nominal and transient wind results and observations from Pioneer Venus, Venus Express, and Magellan, though improvements could be made by incorporating sulfur microphysics.

## 1. INTRODUCTION

Sulfuric acid aerosols make up most of the global cloud deck and accompanying hazes that shroud the surface of Venus (Esposito et al. 1983). As a result, the radiation environment and energy budget at the surface and throughout the atmosphere is strongly affected by the vertical extent, size distribution, and mean optical properties of these particles. These aerosols also serve as a reservoir for sulfur and oxygen, and thus play a major part of the global sulfur oxidation cycle (Mills et al. 2007). Furthermore, recent studies by Zhang et al. (2010; 2012a) have hypothesized that the upper haze layer could provide the source of sulfur oxides above 90 km. Therefore, studying aerosols is a crucial step in understanding the climate and chemistry on Venus.

Observations from the Pioneer Venus atmospheric probes (Knollenberg and Hunten 1980) helped constrain the number density and size distribution of the aerosols in the cloud deck, and revealed the possibility of two size modes with mean radii ~0.2 μm (mode 1) and ~1 μm (mode 2), along with a third, controversial mode with radius ~3.5 μm whose existence has been challenged (Toon et al. 1984). The clouds were also vertically resolved into three distinct regions: the upper cloud, from 58 to 70 km; the middle cloud, from 50 to 58 km; and the lower cloud, from 48 to 50 km. Mode 1 particles have the largest number densities at all altitudes, while modes 2 and 3 particles are relatively more abundant in the middle and lower clouds than in the upper cloud (Knollenberg and Hunten 1980). Both entry probe (Knollenberg and Hunten 1980; Esposito et al. 1983) and remote sensing (Crisp et al. 1989; 1991; Carlson et al. 1993; Grinspoon et al. 1993; Hueso et al. 2008) indicate that the middle and lower clouds are much more variable than the upper cloud. This variability may be associated with strong convective activity within the middle cloud, where downdrafts with amplitudes as large as 3 m $s^{-1}$ and updrafts as large as 1 m $s^{-1}$ were measured *in situ* by the VEGA Balloons (Ingersoll et al. 1987; Crisp et al. 1990).

These observations of the Venus clouds have been interpreted using numerical models that account for transport and/or aerosol microphysics. Toon et al. (1982) showed that sulfur could be present in the upper cloud under low oxygen conditions in sufficient amounts to form mode 1 particles, with mode 2 particles arising from the coagulation of these particles and sulfuric acid droplets. However, they did not model any other interactions between sulfuric acid and the sulfur particles beside coagulation. Krasnopolsky and Pollack (1994), meanwhile, showed that the lower cloud is formed by upwelling and subsequent condensation of sulfuric acid vapor due to the strong gradient in sulfuric acid mixing ratio below the clouds. James et al. (1997) showed that this process is very sensitive to the local eddy diffusion coefficient, and suggested that the variability of the lower and middle clouds was tied to the dynamical motions of the atmosphere in this region. This conclusion was also reached by McGouldrick and Toon (2007); they showed that organized downdrafts from convection and other dynamic processes could produce holes in the clouds. Indeed, observations from Pioneer Venus indicated that this region of the atmosphere has a lapse rate close to adiabatic, with parts of the middle cloud region being superadiabatic (Seiff et al. 1980; Schubert et al. 1980). Imamura and Hashimoto (2001) modeled the entire cloud deck, and reached many of the same conclusions as James et al. (1997) and Krasnopolsky and Pollack (1994) regarding the lower and middle clouds, and Toon et al. (1982) regarding the upper cloud. They also concluded that an upward wind may be necessary in order to reproduce the observations.

The clouds lie below an upper haze (UH), which extends from 70 to 90 km (Mills et al. 2007). In Imamura and Hashimoto's model (2001), small cloud particles are lofted by upward winds out of the top of the model domain, which would place them in this UH. This demonstrates that regional and/or global dynamical processes will lead to some mixing of the haze with the clouds, resulting in variability of the particle populations in the UH, especially if these processes vary with space and time. Though the variability of winds at the clouds-haze boundary has never been measured directly, we do observe the particle population variability. For instance, data from the Pioneer Venus Orbiter Cloud Photopolarimeter (OCPP) revealed latitudinal variations of an order of magnitude in haze optical thickness from the polar region (where it is more abundant) to the tropics, as well as temporal variations on the order of hundreds of days (Kawabata et al. 1980). More recently, Wilquet et al. (2009, 2012) used Venus Express SPICAV/SOIR solar occultation observations to show the existence of bimodality in the size distribution of the UH, with a small mode of radius 0.1-0.3 μm, and a large mode of radius 0.4-1.0 μm. These modes are not to be confused with the aforementioned modes 1, 2, and 3 in the cloud deck, even though they might be physically connected. Interestingly, the mean size of the haze particles as reported by Kawabata et al. from OCPP measurements 30 years earlier (0.23 ± 0.04 μm) lies well within the small mode size range. In addition, Wilquet et al. (2009) find that the extinction of the haze was observed to vary by as much as an order of magnitude in a matter of days. The degree of variability also changed, as observations a few months later (Wilquet et al. 2012) showed variability in the magnitude of the haze extinction of only a factor of two. Time variability of the haze was also observed in infrared images of the Venus southern hemisphere, where the appearance of the haze changed dramatically across tens of degrees of latitude in the span of a few days (Markiewicz et al. 2007). The three studies above also showed that the haze optical depth can exceed unity, making it an active participant in the regulation of solar radiation reaching lower altitudes, and its variability a property that requires better understanding.

However, numerical models with adequate microphysics that include the UH are rare. Yamamoto and Tanaka (1998) and Yamamoto and Takahashi (2006) included the UH in their simulations of aerosol transport via global atmospheric dynamics and reproduced much of the observations satisfactorily. However, the aerosol microphysics in both studies is inadequate due to the lack of a detailed treatment of nucleation.

In this study, we investigate the formation and evolution of the UH and the cloud decks by constructing a one-dimensional (1D) microphysical and vertical transport model that couples the clouds to the haze with a more detailed treatment of the microphysics. We propose two possible causes for the bimodal size distribution and time variability of the haze: (1) the two modes are produced from two separate processes – one mode is derived from the *in situ* nucleation of sulfuric acid onto meteoric dust, a possibility discussed by Turco et al. (1983) for terrestrial atmospheres, and the other mode is made up of cloud particles that have been lofted into the UH via winds and eddy diffusion, and (2) the two modes and the time variability are entirely due to strong transient winds at the cloud tops lofting both mode 1 and mode 2 cloud particles into the UH.

We describe our basic model in section 2, with emphasis on the model attributes unique to our investigation of aerosols in the Venus atmosphere. In section 3 we present our model results, along with comparisons with data from Pioneer Venus and Venus Express. We also discuss our results in the context of physical processes involved in our model. We summarize our work and state our conclusions in section 4.

## 2. MODEL

We use version 3.0 of the Community Aerosol and Radiation Model for Atmospheres

(CARMA) as our base microphysical and vertical transport code. The model is an upgrade from the original CARMA (Turco et al. 1979, Toon et al. 1988) by Bardeen et al. (2008; 2011). We describe our model setup and departures from the base model below, and we refer the reader to Turco et al. (1979), Toon et al. (1988, 1989), and Jacobson et al. (1994) for detailed descriptions of the basic microphysics and vertical transport and English et al. (2011) for the sulfate microphysics in CARMA.

*2.1. Model Setup*

The microphysical and dynamical processes included in the model are the nucleation of liquid sulfuric acid droplets on sulfur and meteoric dust condensation nuclei; the condensational growth, evaporation, and coagulation of these particles; and their transport by sedimentation, advection and diffusion.

Table 1 summarizes the simulation parameters. The model atmosphere extends from 40 to 100 km, covering the altitudes of the cloud deck and UH. This vertical range is split into 300 levels of 200 m thickness each in our model. Our model time step is 10 seconds, and we found that a total simulation time on the order of 2 x $10^8$ seconds, or about 2000 Earth days, was necessary for the model to reach steady state. This is similar to the characteristic vertical diffusion time of the lower clouds as calculated from the eddy diffusion coefficient profile in section 2.4 and far greater than that of the Venus mesosphere (i.e. the altitudes of the upper cloud and upper haze) calculated by Imamura (1997).

In order to cover the size range from meteoric dust to large droplets and represent both volatile and involatile particles, we use two groups of particle bins, each covering the radius range from 1.3 nm to ~30 μm. The lower radius limit corresponds to the size of meteoric dust as described in Kalashnikova et al. (2000), while the upper radius limit mirrors the upper limit of Imamura and Hashimoto's model (2001). The inclusion of multiple bins for involatile particles differs from the approach by Imamura and Hashimoto (2001), which only had the smallest particle size bin allocated to involatile particles.

Figure 1 shows the temperature and pressure profiles used (Seiff et al. 1985), which were fixed in the model. This is a simplification, as episodic increases in the temperatures of the Venus upper mesosphere and lower thermosphere have been known to exist since they were first observed by Clancy and Muhleman (1991). Such temperature increases have also been seen more recently by the Venus Express SPICAV and SOIR instruments (Bertaux et al. 2007) and the VeRa experiment (Tellmann et al. 2009). An increase in temperatures will suppress $H_2SO_4$ aerosol formation and enhance particle evaporation near the top of the domain studied here. These temperature fluctuations will have little direct impact on the particle populations at lower altitudes, but may indicate the presence of substantial changes in the dynamics of the mesosphere, which could affect particle transport.

The production rates of sulfur and sulfuric acid depend on the chemical pathways that lead to their production. Imamura and Hashimoto (2001) used

$$3SO_2 + 2H_2O \rightarrow S + 2H_2SO_4 \qquad (1a)$$

(Yung and DeMore 1982; Krasnopolsky and Parshev 1983) as their primary reaction. However, Yung and DeMore (1982) suggests that the main scheme for production of sulfuric acid is actually

$$SO_2 + \frac{1}{2}O_2 + H_2O \rightarrow H_2SO_4 \qquad (1b)$$

where the sulfur produced during the reaction is converted to SO via reaction with $O_2$, though their model still shows a net production of S. Furthermore, equation 1a is derived from Yung and DeMore (1982)'s Model A, which is more appropriate for a Venus early in its evolution than the current Venus. Meanwhile, Krasnopolsky and Parshev (1983) note that the reaction that would normally generate S, SO + SO, could also go on to produce $S_2O$ instead. All of these considerations suggest that the production rate of S is likely to be lower than half the production rate of sulfuric acid, as suggested by equation 1a. On the other hand, Yung and DeMore (1982) also showed that polysulfur can be produced and that sulfuric acid production (equation 1b) can be suppressed via $(SO)_2$ dimer chemistry, while Toon et al. (1982) suggested that the primary reaction can switch between equations 1a to 1b depending on the local $O_2$ content, which may be variable. Therefore, it is uncertain what the sulfur production rate actually is. For simplicity, we will use equation 1a as a basis for the production of sulfuric acid and sulfur, but we will also test the effect of decreasing the sulfur production rate (section 3.3).

We begin each model run with no model-relevant species in the model box, e.g. no sulfuric acid vapor or condensation nuclei of any kind. As each model run progresses, mass is injected into the model atmosphere in the form of sulfuric acid vapor and involatile condensation nuclei. The latter is split into two populations, one corresponding to photochemical products (sulfur), and one corresponding to meteoric dust. Here, we assume a density of 1.9 g $cm^{-3}$ for the condensation nuclei, as an average between the density of sulfur (1.8 g $cm^{-3}$, Imamura and Hashimoto 2001) and meteoric dust (2.0 g $cm^{-3}$, Hunten et al. 1980). We use the same production profiles of sulfuric acid vapor and photochemical condensation nuclei as Imamura and Hashimoto (2001) for the production rates $P_{H2SO4}$ and $P_{CN}$, respectively:

$$P_{H_2SO_4} = \Phi_p g(z) \ \ \mathrm{cm^{-3}s^{-1}} \tag{2a}$$

$$P_{CN} = \frac{1}{2}\Phi_p g(z)\left(\frac{\rho_{CN}}{M_s}\frac{4}{3}\pi r_{CN}^3\right)^{-1} \ \ \mathrm{cm^{-3}s^{-1}} \tag{2b}$$

where $\Phi_p$ is the column-integrated production rate of sulfuric acid vapor; the function g(z) is a Gaussian with a peak at 61 km altitude and full-width-half-max of 2 km such that

$$\int_0^\infty g(z)\, dz = 1 \quad (2c)$$

$\rho_{CN}$ is the density of the condensation nuclei, 1.9 g $cm^{-3}$; $r_{CN}$ is the radius of the condensation nuclei; and $M_s$ is the mass of a sulfur atom, 5.34 x $10^{-23}$ g. Our results showed that agreement between model and data was best if $\Phi_p$ was decreased from Imamura and Hashimoto's (2001) nominal value of $10^{12}$ $cm^{-2}$ $s^{-1}$ to 6 x $10^{11}$ $cm^{-2}$ $s^{-1}$, consistent with the suppression of the sulfuric acid production rate discussed in Yung and DeMore (1982). Our nominal production profiles are plotted in Figure 2.

The exact mechanics of how sulfuric acid nucleates onto condensation nuclei is not well understood and this is made worse by the complexities of the chemistry in the Venus atmosphere. As previously stated, we assume that the photochemical condensation nuclei are made of sulfur, similar to the strategy of Imamura and Hashimoto (2001). However, this is a simplification, as sulfur would likely exist in the form of polysulfur in the Venus clouds. Polysulfur would also undergo the processes of nucleation, condensation, and evaporation similar to sulfuric acid aerosols (Toon et al. 1982), with the only difference being that the polysulfur aerosols would likely be solid at the temperatures of the upper cloud (Lyons 2008; Zhang et al. 2012a) and therefore would not coagulate as efficiently as liquid aerosol droplets. Furthermore, as sulfuric acid cannot actually wet sulfur (Young 1983), polysulfur cannot act as condensation nuclei in the sense that they form cores that are completely encased within a layer of sulfuric acid. Instead, it is more likely that sulfuric acid will only condense on a fraction of the total surface of a polysulfur particle, and it is this small “drop” of condensed acid that then acts as the nucleation site for more sulfuric acid. Thus, the actual particles would be made up of a polysulfur particle stuck to the side of a droplet of sulfuric acid, with part of the polysulfur particle exposed to the atmosphere. As Young (1983) elucidates, this has the effect of decreasing the efficiency of coagulation in the growth of these sulfuric acid aerosols, as now part of the surface is covered by polysulfur and will not be able to participate in coagulation. In addition, as coagulation occurs, more of the sulfuric acid particle's surface area will be covered by polysulfur (from particles it coagulated with), further decreasing its coagulation rate. This has the effect of eventually stopping coagulation altogether when the particle reaches a radius of ~10 μm. While our model does not distinguish whether the polysulfur "core" is within the sulfuric acid or attached to its side, it does assume, for coagulation, that the entire surface is "available". We will discuss the effect of this in section 3.1.

Further complicating the picture is the process opposite to the one we are modeling: the nucleation of polysulfur onto sulfuric acid particles (Young 1983; Lyons 2008). Whether one process dominates the other is determined by which particles homogeneously nucleates, i.e. which one appears "first" to act as the condensation nuclei for the other. In this case, as the saturation vapor pressure of most species of $S_x$ is an order of magnitude or more lower than that of sulfuric acid in the cloud top region (Young 1983; Lyons 2008), we can safely assume that polysulfur will homogeneously nucleate before sulfuric acid does so as to act as its condensation nuclei and that we may ignore the opposite process. However, future studies should take into account the aerosol physics of the polysulfur nuclei to investigate its effects on the cloud distribution.

In our model, nucleation of sulfuric acid vapor onto the polysulfur condensation nuclei occurs when the ambient sulfuric acid vapor concentration reaches a critical supersaturation. This is determined by the sulfuric acid weight percent of the liquid that would form upon condensation (see section 2.2) and the curvature of the surface on which the vapor will condense. That is, we assume that the accommodation coefficient is unity. The curvature effect is determined by the Kelvin equation (Seinfeld and Pandis 2006):

$$Ln\frac{p}{p_o} = \frac{2\gamma M}{\rho r R T}. \quad (3)$$

For our case, $p_o$ is the saturation vapor pressure over a flat surface; p is the saturation vapor pressure over a surface with curvature r; γ, M, and ρ are the surface tension, molar mass, and density of sulfuric acid, respectively; R is the gas constant; and T is the temperature. For reasonable values of these parameters corresponding to ~61 km in the Venus atmosphere, where we use the parameterization of Mills (1996) for the surface tension, we find that the actual saturation vapor pressure over condensation nuclei becomes an order of magnitude greater than the saturation vapor pressure over a flat surface when the radius of the particle is ~2 nm. However, given equation 2a, it will take on the order of hours for the sulfuric acid to build to such concentrations, during which time the polysulfur condensation nuclei will also be growing, thus increasing its radius of curvature and decreasing the saturation vapor pressure over its surface. Therefore, there may exist an equilibrium point where the sulfuric acid concentration is just high enough to begin nucleation onto the growing polysulfur particles. The calculation of this equilibrium point is beyond the scope of this work, however, as our knowledge of the kinetics of sulfur reactions is poor and thus we have ignored all microphysics (nucleation, condensation, coagulation) associated with the condensation nuclei. Therefore, we will make a simplification and assign a radius of ~10 nm to the polysulfur condensation nuclei that are injected into the model. Toon et al. (1982) simulated the microphysics of sulfur particles, but did not take into account the effects of background sulfuric acid vapor or nucleation of said vapor on the sulfur particles. It would be ideal to combine these processes to create a more complete picture of the Venus clouds and hazes.

We adopt similar lower boundary conditions as those of Imamura and Hashimoto (2001),

where involatile (sulfur) particles of size ~0.17 μm (mode 1) are fixed to have a number density of 40 $cm^{-3}$ in accordance with LCPS data (Knollenberg and Hunten 1980). We set the mixing ratio of $H_2SO_4$ to be 3 ppm at the lower boundary, within the 0-4 ppm estimates from analysis of Magellan radio occultation observations by Kolodner and Steffes (1998). We adopt a zero flux boundary condition for the top boundary, as we assume that no particles or $H_2SO_4$ vapor escape the mesosphere above 100 km.

*2.2. Thermodynamics of $H_2SO_4$*

Of particular importance in this model is the treatment of certain thermodynamic properties of $H_2SO_4$, such as its saturation vapor pressure and surface tension. Both of these quantities control whether a sulfuric acid droplet is growing by condensation or evaporating.

The saturation vapor pressure $p_{H2SO4}$, in units of atm, is calculated via the equation of Ayers et al. (1980), modified by Kulmala and Laaksonen (1990):

$$\ln p_{H_2SO_4} = \ln p^0_{H_2SO_4} + 10156\left[-\frac{1}{T} + \frac{1}{T_0} + \frac{0.38}{T_c - T_0}\left(1 + \ln\frac{T_0}{T} - \frac{T_0}{T}\right)\right] - \frac{H}{RT} \tag{4a}$$

where T is temperature, R is the universal gas constant, $T_o$ = 340 K is a reference temperature, $T_c$ = 905 K is the critical temperature, $p^0_{H2SO4}$ is a reference pressure given by:

$$\ln p^0_{H_2SO_4} = -\frac{10156}{T_0} + 16.259 \tag{4b}$$

and H is the enthalpy associated with the mixing of water and sulfuric acid, given by the parameterization of Giauque (1959):

$$H = 4.184 \times 10^7\left[23624.8 - \frac{1.14208\times10^8}{4798.69 + \left(W_{H_2SO_4} - 105.315\right)^2}\right] \quad erg\ mol^{-1} \tag{4c}$$

where $W_{H2SO4}$ is the weight percentage of $H_2SO_4$ in the aerosol droplet calculated from Tabazadeh et al. (1997) as a parameterization to temperature and background water vapor concentration.

The surface tension is derived from data collected by Sabinina and Turpugow (1935) parameterized linearly with respect to temperature by Mills (1996). The value of the surface tension at ~61 km is given in Table 1.

*2.3. Water Vapor Profile*

The base CARMA model does not treat the exchange of water between the background water vapor concentration and the water taken up in the sulfuric acid cloud particles as they grow. Instead, the weight percentage of water and sulfuric acid is always such that there is equilibrium between the particle and the background water vapor concentration. In other words, the model assumes the equilibrium aerosol growth regime (Zhang et al. 2012b). However, this may not necessary be true given the low water vapor concentration in the Venus atmosphere (Ignatiev et al. 1997). Therefore, we will assume an equilibrium background water vapor profile that is fixed in time. Figure 3 shows this profile. Above 60 km, we assume a water vapor concentration of ~1 ppm in accordance with observations by Bertaux et al. (2007). Below 60 km, we model our profile after Model A of Ignatiev et al. (1997) and data from Venera 11, 13, and 14 (also taken from Ignatiev et al. 1997). The profile is empirically determined using an arctangent function to connect the region above 60 km to the region below, and a Gaussian function to take into account the water vapor concentration maximum at ~53 km in Model A of Ignatiev et al. (1997). The formula for the water vapor concentration (partial pressure), $P_{H2O}$, is then

$$P_{H2O} = 6 \times exp\left[-\left(\frac{z-53}{8}\right)^2\right] - 14.5\left(\frac{2}{\pi}\right) tan^{-1}(0.5z - 32.5) + 15.5 \ \ ppm \qquad (5)$$

where z is in kilometers.

*2.4 Eddy Diffusion*

The eddy diffusion coefficient profile is shown in Figure 4. The values between 40 and 70 km altitude are approximated from Imamura and Hashimoto (2001) by fitting a function consisting of the sum of an exponential and a Gaussian function. The large increase in eddy diffusion coefficient at ~53 km simulates the convective overturning present in the middle cloud as inferred from Schubert et al. (1980) and measured *in situ* by the VEGA Balloons (Ingersoll et al. 1987; Crisp et al. 1990). The eddy diffusion coefficient above 70 km is approximated as a Gaussian from Fig. 11 of Krasnopolsky (1983), which itself is generated from continuity arguments with respect to the aerosol distribution observed in this region at the time. The empirical formula of the eddy diffusion coefficient, $K_{zz}$, as a function of altitude z in kilometers above 40 km is then:

$$K_{zz} = 10^{\frac{4z-160}{38.55}} + 2500000\left\{exp\left[-\left(\frac{z-52.5}{1.201}\right)^2\right] + exp\left[-\left(\frac{z-100}{12.01}\right)^2\right]\right\} \quad cm^2\, s^{-1} \tag{6}$$

To implement eddy diffusion in CARMA 3.0, we adopt similar numerical methods used by the model to implement Brownian diffusion, except we replace the density of the species by its mixing ratio. The upward and downward velocities of eddy diffusion of species j, $v^j_u$ and $v^j_d$, respectively, are then given by:

$$v_u^j = \ln\left(\frac{f_i^j}{f_{i-1}^j}\right)\frac{K_{zz}}{dz}\frac{f_{i-1}^j}{f_i^j - f_{i-1}^j} \quad cm\, s^{-1} \tag{7a}$$

$$v_d^j = \ln\left(\frac{f_i^j}{f_{i-1}^j}\right)\frac{K_{zz}}{dz}\frac{f_i^j}{f_i^j - f_{i-1}^j} \quad cm\, s^{-1} \tag{7b}$$

where dz is the thickness of an atmospheric layer (200 m in our model) and $f^j_i$ is the mixing ratio of species j in the ith layer. We can see that equations 7a and 7b are diffusion velocities by considering the continuous limit, where $Ln(f_i/f_{i-1}) = Ln(f_i) - Ln(f_{i-1}) \sim dLnf = df/f$, and $f_i - f_{i-1} = df$; the two df's and f's then cancel, resulting in $K_{zz}/dz$, which is an eddy diffusion velocity scale. This velocity is modulated by the mixing ratio gradient in our discrete case (i.e. $Ln(f_i/f_{i-1})$ does not cancel with $f_i/(f_i - f_{i-1})$) such that large gradients lead to large velocities. The natural log prevents numerical instabilities in the event the denominator becomes too small.

*2.5. Meteoric Dust*

Turco et al. (1983) discussed the properties of meteoric dust in the Venus atmosphere and concluded that it is similar to meteoric dust in the atmosphere of Earth and could act as condensation nuclei to water vapor, forming thin ice hazes. We propose that meteoric dust could also serve as condensation nuclei to sulfuric acid vapor, as its saturation vapor pressure is extremely low at the altitude of the UH, on the order of $10^{-19}$ mbars for pure sulfuric acid, and $10^{-31}$ mbars for a water-sulfuric acid mixture with 75 wt% sulfuric acid (Kulmala and Laaksonen 1990) that is typical of the UH (Kawabata et al. 1980). Thus, any sulfuric acid vapor that is lofted into the UH by diffusion or winds could potentially condense on the meteoric dust present in this region.

The Kelvin effect may play a large role in limiting the efficiency of meteoric dust as condensation nuclei due to their small size. However, if we use the appropriate values for sulfuric acid in the UH and a typical condensation nuclei size of 1.3 nm (Kalashnikova et al. 2000), then equation 3 yields an approximate increase of 2.4 orders of magnitude in the saturation vapor pressure. This gives a resulting saturation vapor pressure of ~ $10^{-29}$ - $10^{-28}$ mbars, which is far less than recent upper limits on the abundance of $H_2SO_4$ in the UH, e.g. 3 ppb, or about $3x10^{-9}$ mbar at 80 km, from Sandor et al. (2012). Therefore, meteoric dust should act as very efficient condensation nuclei as long as the actual $H_2SO_4$ mixing ratio is not significantly lower than this upper limit, even if the Kelvin effect is considered.

Meteoric dust is treated in the same way in our model as the sulfur condensation nuclei. However, it is clear that meteoric dust, which is typically made of silicates (Hunten et al. 1980), may react differently to sulfuric acid than sulfur. For example, Saunders et al. (2012) showed that silicates dissolve in sulfuric acid on a timescale of an Earth week at the temperatures of the cloud tops. However, once the sulfuric acid droplet becomes much larger than the nucleus it condensed on, the dissolution of said nucleus should have very little effect on the rest of the particle.

The production profile of meteoric dust we use in our model is shown in Figure 5 as an empirical approximation of the profile calculated by Kalashnikova et al. (2000). All meteoric dust particles are assumed to have a radius of 1.3 nm. We have shifted the profile maximum from 87 km to 83 km in order to match the maximum in the small mode curve in Fig. 9 of Wilquet et al. (2009). The parameterization of the profile is given by:

$$P_{md} = \begin{cases} 5 \times 10^{-3} e^{-\left(\frac{z-83}{2.402}\right)^2} & z \leq 83 \\ 5 \times 10^{-3} e^{-\left(\frac{z-83}{8.407}\right)^2} & z \geq 83 \end{cases} \tag{8}$$

where z has units of kilometers and $P_{md}$ has units of $cm^{-3}$ $s^{-1}$. Though this profile was applied to Earth only, we will assume that it is also applicable to Venus due to the two planets' similarities. However, if atmospheric density is the determining factor of the altitude of peak meteoric dust ablation (e.g. Gadsden 1968) then our profile may be too low in the atmosphere. This is due to the fact that the altitude on Venus with the same atmospheric density as that of the peak ablation altitude on Earth is actually ~110 km.

*2.6. Winds*

In addition to testing the effects of the nucleation of sulfuric acid droplets on meteoric dust on the UH particle size distribution, we will also test the effects of transient gusts at the cloud tops. These are a separate set of simulations from the nominal runs described in section 2.1, though they do use the results of the nominal runs as initial conditions. Figure 6 shows the wind profile we use to test the effects of transient upward winds on the number density and size distribution of the cloud and haze aerosols. The wind beneath 70 km is a constant flux wind similar to that of Imamura and Hashimoto (2001) but increased in strength by two orders of magnitude to simulate a gust as opposed to a branch of the global circulation:

$$w = \frac{8.0\times10^{-3}}{\rho} \quad cm\ s^{-1} \tag{9}$$

where w is upward wind speed and ρ is atmospheric density, both in cgs units. In order to adhere to our top boundary condition and simulate turning over of the wind currents, we allow the upward wind to fall off linearly above 70 km so that it vanishes at 75 km. This is consistent with the strong static stability above that altitude.

Recent Venus Express observations of the Venus southern polar vortex at the altitude of the upper cloud (63 – 69 km) show divergent and convergent circulations that imply vertical velocities of ~0.2 m $s^{-1}$ (Garate-Lopez et al. 2013), roughly consistent with the lower limits of our upper cloud wind velocities, though most of the vertical motions were downward rather than upward. In addition, convective cells at the cloud tops have been observed at and downwind of the Venus subsolar point (Belton et al. 1976; Titov et al. 2012). These have been interpreted using 2D convection simulations as the incursion of the convective region in the middle cloud into the stably stratified upper cloud (Baker et al. 1998; 1999), though some recent observations suggest that the convective layer is thin enough that it may only occupy the upper cloud itself (Markiewicz et al. 2007). The results of Baker et al. (1998) suggest that gravity waves originating from the enlarged convective region could produce vertical velocities of 1-2 m $s^{-1}$ at and above 60 km, matching our wind velocities at those altitudes.

Nevertheless, there are clear differences between our vertical advection scheme and the actual advection processes in the Venus clouds. For instance, a realistic treatment of the gust "turning over" above 70 km requires horizontal transport, which is beyond the scope of this study. Thus, for the purpose of this work we will only examine qualitatively the effects of such a gust on the cloud profile, such as the formation of a detached haze due to upwelled cloud particles, its destruction due to sedimentation and diffusion, the time scales involved, and the changes in particle size distributions. In order to better evaluate these effects, we will magnify them by considering gusts that last for 5 x $10^4$ seconds, or ~14 Earth hours. Gusts in the polar vortex may last as long, though as Venus Express orbits Venus on the same time scale (~1 Earth day), the actual duration is uncertain. Winds caused by subsolar point convection may also be able to last 14 hours, as convective features were seen up to 50° latitude away from the equator (Titov et al. 2012). This should also apply to longitude, and a distance of 50° longitude away from the subsolar point can be covered by the ~100 m $s^{-1}$ planet-encircling zonal winds at the cloud tops (Schubert et al. 1980) in about 14 hours. In other words, each parcel of air could have 14 hours during which gusts arising from subsolar convection can disrupt the aerosol particle distributions therein before the convective cells give way to stably stratified regions. Nevertheless, not only is the wind unlikely to be constant given the turbulent nature of its origin, but Wilquet et al. (2009) observed the haze at ~70° N, a full 20° latitude away from the convective region. In short, we are not looking for exact agreement between our particle number density and size distributions and one that is retrieved from observations, but rather whether we can qualitatively reproduce the chaotic behavior of the UH.

## 3. RESULTS AND DISCUSSION

Our steady state results do not show a stable equilibrium distribution, but rather a quasi-periodically varying distribution. Thus, we will proceed to compare to observations our results from a single time step near the end of our nominal model run that best match them (section 3.1), and then describe the mechanisms that could allow for such dynamics to occur despite a constant background atmospheric state (section 3.2). In section 3.3 we discuss the effect of decreasing the sulfur production rate, and in section 3.4 we will describe our transient wind results.

*3.1 Nominal Results*

Figure 7 shows the number density predictions from our model for particles with radii greater than ~0.1 μm, which we choose as the lower size boundary of mode 1/small mode particles (Wilquet et al. 2009). We see that it is largely consistent with LCPS cloud data (Knollenberg and Hunten 1980), though the middle cloud is overestimated by a factor of 3 and the lower cloud is underestimated by a factor of < 2. These differences are well within the range of variability seen at these levels by the Pioneer Venus entry probes (Esposito et al. 1983), or in near infrared observations of the Venus night side (Crisp et al. 1990; Grinspoon et al. 1993). Figure 8 shows the size distributions of the cloud and haze particles at various altitudes. It reproduces both the high number density of mode 1 particles in the upper cloud and the distinct multi-modal nature of the middle and lower clouds' particle size distributions as seen by Pioneer Venus (Knollenberg and Hunten 1980), including mode 1 at ~0.2 μm, mode 2 at ~1.4 μm, and mode 3 (near the cloud base) at ~3.5 μm. Comparison of this size distribution with Pioneer Venus observations at 54.2 km (Knollenberg and Hunten 1980) shows clear agreement between the model mode radii and that of the data, though a discrepancy exists in the mode abundances – there is an order of magnitude less mode 3 particles and ~3 times more mode 2 particles in our model than in the observations. These discrepancies could be caused by the lack of transient gusts in the middle and lower clouds of our steady state model, which Imamura and Hashimoto (2001) showed could be used to produce a multi-modal structure in their model. Physically, vertical gusts such as those detected in the middle cloud by the VEGA Balloons (Crisp et al. 1990) would both aid in the growth of mode 2 particles into mode 3 particles by introducing "fresh" sulfuric acid from below (updrafts) and in the depletion of the middle cloud by downwelling the cloud particles (downdrafts); the depletion of the middle cloud is also a natural consequence of the growth of particles, as these larger particles would have a faster sedimentation time. Therefore, a series of vertical updrafts and downdrafts could lead to fewer mode 2 particles, more mode 3 particles, and fewer particles in the middle cloud, resulting in better agreement with data.

We assume that mode 3 particles are just larger versions of mode 2 particles, i.e. they are liquid sulfuric acid droplets with a solid polysulfur component, but this need not be the case. In fact, Knollenberg and Hunten (1980) discussed the possibility that mode 3 is made up of solid, crystalline particles. On the other hand, Toon et al. (1984) suggested that mode 3 is in fact a large particle tail of mode 2; our results support this latter interpretation, as the mode 3 particle peak blends into the mode 2 peak somewhere between 54 and 58 km (figure 8). In light of the results of Young (1983), however, we must consider the effect of solid polysulfur "patches" on the surface of our sulfuric acid particles, which may make them seem like solid, crystalline particles if the polysulfur coverage is high enough. As previously stated, the effect of these patches is the decrease in the efficiency of coagulation in the removal of smaller particles, creating a smaller number of larger particles. This could explain our underestimation of the lower cloud particle number density, though it would also further overestimate our middle cloud particle number density. Thus, a combination of decreased coagulation efficiency in larger particles and the addition of middle cloud vertical gusts may be necessary to improve our model's agreement with the Pioneer Venus Sounder Probe observations.

In the UH, our model predicts a steady state size distribution that is roughly mono-modal, with a small "bump" at radii > 1 μm corresponding to upwelled mode 2 particles that is largely insignificant compared to the rest of the distribution. The lack of distinct bimodality despite the two different sources of particles is likely due to the effects of coagulation. In Figure 9, we follow the time evolution of the upper haze at an altitude of 84 km during the first ~4 months of the model run, long before steady state is reached. The green "bar" at 0.01 μm is caused by the artificial injection of 10 nm photochemical condensation nuclei into the model domain and should be ignored. Immediately after the start of the simulation, a haze population emerges with particle radii < 10 nm. Such small particles could have only resulted from the nucleation of sulfuric acid onto meteoric dust and subsequent condensational growth. Cloud droplets reach 84 km one week later due to upward diffusion, which is followed by the disappearance of the original meteoric-dust-derived haze population about a month afterwards. This suggests a coagulation timescale of about a month to decrease the original population by about 2-3 orders of magnitude, if indeed coagulation is the cause of this result. We can test whether this is physically viable using the simplified solution to the discrete coagulation equation N(t):

$$N(t) = \frac{N_0}{1+(t/\tau_c)} \tag{10a}$$

where

$$\tau_c = \frac{2}{KN_0} \tag{10b}$$

is a coagulation timescale. Here, $N_0$ is the original number density, and K is the coagulation kernel. For simplicity we use the continuum regime kernel, as particle number density is small, and assume that the particle radii $R_2 >> R_1$:

$$K \sim \frac{2kT}{3\mu}\frac{R_2}{R_1} \tag{11}$$

(Seinfeld and Pandis 2006). For a Boltzmann's constant $k = 1.38 \times 10^{-16}$ erg $K^{-1}$, temperature T ~ 160 K, atmospheric viscosity $\mu \sim 8 \times 10^{-5}$ g $cm^{-1}$ $s^{-1}$ and particle radii $R_1$ = 1 nm and $R_2$ = 100 nm, we get a coagulation time needed to reduce the number density by ~2-3 orders of magnitude of a few months, similar to our model results. Therefore we can conclude that the UH is likely a combination of upwelled cloud particles and particles nucleated *in situ* on meteoric dust, but that the latter population has “fused” with the former, resulting in a mono-modal size distribution in steady state. Thus, two separate sources of particles – one upwelled from below, and one nucleated *in situ* cannot explain the bimodality detected by Kawabata et al. (1980) and Wilquet et al. (2009). However, our peak model cloud top haze particle radius, ~0.3 μm at 70 km, is fairly close to the average haze particle radius originally detected by Kawabata et al. (1980) (0.23 ± 0.04 μm). Furthermore, our UH size distribution covers the size range of both the small and the large mode detected by Wilquet et al. (2009). Therefore, we propose that our results represent both modes, and that something else is responsible for the splitting of the particle population into two distinct modes. Figure 7 shows a comparison between the model number density and the sum of the number densities of the two modes detected by Wilquet et al. (2009). The agreement between model and data is satisfactory up to 80 km, above which the model underestimates the number density by about half an order of magnitude.

Alternatively, the meteoric dust production profile could be at a much higher altitude (see section 2.5) such that they are above the upwelled cloud particles; the sulfuric acid droplets that nucleate onto these meteoric dust particles would then have time to grow before they sediment into the altitudes dominated by cloud droplets, with which they will coagulate as before. In this scenario, the small mode would consist of both the upwelled cloud particles and the sedimenting particles that nucleated *in situ* onto meteoric dust, while the larger mode would be made up of the products of small mode coagulation. However, the resulting smaller $R_2/R_1$ ratio (equation 11) would increase the coagulation time scale and decrease the amount of large mode particles that can be produced. Ultimately, the particles that nucleate from meteoric dust may be inconsequential due to the lower flux of meteoric dust as compared to the flux of sulfuric acid vapor and sulfur.

Figure 10 shows a comparison between the predicted mixing ratio of sulfuric acid vapor and the observed mixing ratio from Magellan radio occultation data as analyzed by Kolodner and Steffes (1998). Though the large dispersion in the data from 0-6 ppm allows for a wide variety of results to "fit" it, the physically relevant results are likely those that exhibit a local sulfuric acid maximum below the clouds, and which also fit the nonzero data points. Indeed, our model results show a satisfactory fit to the data points. While the bottom boundary condition fixes the vapor concentration to 3 ppm at 40 km, the vapor concentration peak of ~4.5 ppm at the cloud base is entirely due to sedimenting particles evaporating and depositing their sulfuric acid vapor at that altitude. The peak value reflects a balance between the rate of vapor deposition and the upward and downward vapor diffusion, and also imposes a constraint on the altitude of the cloud base, below which any sulfuric acid droplets will be evaporating. Our results show that the cloud base is at ~47 km, consistent with the model results of Krasnopolsky and Pollack (1994). Above the cloud base, the vapor profile largely follows the saturation vapor pressure curve except at (1) 61 km, where sulfuric acid vapor is photochemically produced, and (2) above 80 km. This latter deviation may be caused by numerical instabilities caused by the low saturation vapor pressure ($\sim 10^{-31}$ mbars) or the phase properties of sulfuric acid and water solutions in this region (McGouldrick et al. 2011).

Figure 11 shows number density as a function of both altitude and particle radius, while figure 12 shows the sulfuric acid and particle mass fluxes for the middle cloud (top) and the other altitudes (bottom); together they give a summary of the processes occurring in the clouds and UH of Venus. The production of involatile sulfur condensation nuclei causes the nucleation and condensational growth of liquid sulfuric acid droplets at 61 km, resulting in a high number density of mode 1 particles. These particles then diffuse upwards and sediment downwards. The positive divergence in the particle flux (in units of mass equivalent to $10^{12}$ sulfuric acid molecules per unit area per second, where each molecule has mass ~1.6 x $10^{-22}$ g) in this region is clearly shown by the positive slope at ~61 km in the lower panel of Figure 12 – particles higher up has a smaller downward flux (i.e. a less negative flux) than the particles lower down. There is no corresponding slope in the vapor flux curve as all the vapor is condensing onto the particles. In the UH, the upward diffusion of particles appears to balance the sedimentation, leading to a near zero particle flux; the sulfuric acid vapor flux is also very close to zero, but it is likely due to the small amount of vapor at these altitudes.

Below 61 km, the vigorous convection in the middle cloud drives the large upward flux of sulfuric acid vapor (top panel of figure 12), resulting in enhanced production of mode 2 particles as well as a tail of even larger particles resulting from coagulation. These particles, like the upper cloud particles, are transported downwards by sedimentation and upwards by diffusion. The latter process leads to a small large particle bump in the UH. These particles evaporate upon reaching the cloud base, leading to the regeneration of mode 1 particles and the deposition of sulfuric acid vapor beneath the clouds. The regenerated mode 1 is larger than the mode 1 of the upper cloud due to the coagulation of the droplets higher up in the clouds – the cores of these droplets are in effect added together, leading to the generation of larger involatile particles as the larger cores are exposed upon droplet evaporation. This may not be what actually happens if the "cores" are instead patches of polysulfur decorating the outside of the sulfuric acid droplets, as per the "gumdrop" model of Young (1983). In this case, the "cores" would not be added together and would remain the same size as their upper cloud counterparts upon the evaporation of the sulfuric acid. Meanwhile, the tail of large particles forms a distinct third mode upon evaporation due to its slower evaporation rate as compared to that of the smaller mode 2 particles, which is a product of the Kelvin effect (equation 3).

Below the clouds, the sulfuric acid vapor exhibits a negative flux as discussed previously, while the particle flux is similarly negative, as the particles are sedimenting out of the bottom of the model domain. The absolute value of the sum of the two fluxes at the bottom of the model domain is ~ 9 x $10^{11}$ $cm^{-2}$ $s^{-1}$, slightly higher than the sum of the input flux of meteoric dust (~5 x $10^5$ $cm^{-2}$ $s^{-1}$), photochemical condensation nuclei (~$10^{11}$ $cm^{-2}$ $s^{-1}$), and sulfuric acid vapor (6 x $10^{11}$ $cm^{-2}$ $s^{-1}$). This is due to the quasi-periodically varying nature of our results, i.e. we have examined our simulated Venus cloud-haze system at a time when the flux out of the lower model boundary is higher than what it should be for a steady state (~6.835 x $10^{11}$ $cm^{-2}$ $s^{-1}$).

*3.2. Periodic Behavior and Precipitation on Venus*

Figure 13 shows the time evolution of the particle size distribution at various altitudes. Above the clouds, the quasi-periodic variations are very small, with amplitudes of no more than 10%. In the upper cloud, the variations are larger with small hints of quasi-periodicity in the mode 2 particle abundances apparent, but the size distribution is dominated by mode 1 particles, which are fairly stable. In contrast, the middle cloud shows variations in particle abundances of several orders of magnitude. All three modes appear to grow in radii with time and are subsequently replaced with smaller particles on a time scale of ~6 Earth months. In particular, mode 1 slowly grows to larger radii via coagulation and condensation until reaching ~0.25 µm, where there is a rapid growth of particles to mode 2. This actually occurs at several points during the slow migration of mode 1 particles to larger radii, as the intense eddy diffusion at 54 km will allow any particles large enough to overcome the Kelvin barrier to grow rapidly. The upward diffusion of mode 1 particles from below then "resets" the mean mode 1 particle radius back to

~0.2 μm. Mode 2 then also grow slowly with time before growing somewhat rapidly into mode 3 particles, which then proceed to sediment into lower altitudes. The increase in mode 2 particle growth rate occurs at nearly the same time as the rapid growth of mode 1 particles into mode 2 particles, possibly due to the coagulation of the "new" mode 2 particles among themselves and with the "old" mode 2 particles, leading to the creation of the larger mode 3 particles.

Figure 13 also shows that, in comparison with the nominal/best-fit results in section 3.1, the "usual state" of the Venus clouds is actually not those observed by the LCPS – for the majority of the time, mode 2 has a mean radius of ~1.2 μm instead of the observed 1.4 μm (Knollenberg and Hunten 1980), and it only grows to such a size right before the emergence of a "new" mode 2.

This oscillation is apparently perpetuated by the growth of mode 1 particles into sizes capable of overcoming the Kelvin Barrier, as that leads to the perturbation of the established mode 2 particles which in turn leads to their growth into mode 3 particles. The growth of mode 1 particles can occur via coagulation among themselves, though it is more likely that the majority of coagulation events are with the large number of smaller particles we use in our model as photochemical condensation nuclei and which grew from those nuclei but have not yet reached the sizes of the mode 1 particles, as coagulation between particles of vastly difference sizes is faster than coagulation between similarly sized particles (Seinfeld and Pandis 2006). As such, since each coagulation event would only add a very small amount of mass to each mode 1 particle, the growth would be gradual but steady, matching what we see in our model results. It is interesting to note that this does not happen when we use a larger particle as our photochemical condensation nuclei (e.g. Imamura and Hashimoto (2001)), as coagulation would be much slower and "random" in time. These variations also do not happen to such an extent in the UH, as both the coagulation kernel (equation 11) and condensational growth rate are much lower there.

The quasi-periodic variability of the middle cloud leads to a quasi-periodic sedimentation flux for the largest particles, which leads to the larger-than-expected bottom boundary flux in figure 11. Figure 14 shows the time evolution of the sulfuric acid vapor and particle flux out of the bottom boundary of the model domain, expressed in the same units as figure 12. The average of these oscillations, $6.765 \times 10^{11}$ $cm^{-2}$ $s^{-1}$ matches the injection rate of mass into the model domain, $6.835 \times 10^{11}$ $cm^{-2}$ $s^{-1}$ fairly well, indicating that we are indeed at an equilibrium state. Finally, we note that the particle and vapor flux oscillations are in phase with each other, and that the latter has greater amplitude than the former. This is explained by the evaporation of the particles below the cloud deck at ~47 km, such that the sulfuric acid variations from the middle cloud are mostly expressed in the resulting vapor.

The possibility of such long term, quasi-periodic particle size variations provide tantalizing hints towards possible sulfuric acid "rain" below the cloud decks of Venus, though we use the term loosely as the downward mass fluxes associated with the "rain" in our model results are far smaller than that of typical rain events on Earth. Nonetheless, it would be interesting to search for evidence of their existence in VIRTIS observations of the Venus night side, which are sensitive to such variations in particle population.

*3.3 Variations in the Sulfur Production Rate*

As discussed in section 2.1, the production rate of sulfur in the Venus atmosphere may be much lower than half the production rate of sulfuric acid as implied by equation 1a due to reactions of S with $O_2$ and the formation of $S_2O$ instead of $SO_2$ and S from the reaction SO + SO. To address the effect decreasing the sulfur production would have on our results we rerun our model with both one order of magnitude less sulfur produced and two orders of magnitude less sulfur produced. Figure 15 shows our results in terms of the number density (top) and size distribution at 54 km in the middle cloud (bottom). Decreasing the sulfur condensation nuclei production rate decreases the number of cloud droplets produced at all altitudes, but also increases the average size of the droplets, since there are fewer nuclei for the same amount of gas to condense on. Mode 1 particle abundances also decrease significantly in the middle cloud when sulfur production is reduced, as the relative fraction of mode 1 particles that grow to mode 2 particles is now much larger. The disagreement between the observed size distribution and the results of the reduced sulfur cases suggest that either (1) the primary reaction in the production of sulfur is indeed reaction 1a (section 2.1), and that $O_2$ is relatively scarce in the upper cloud, or (2) Pioneer Venus took data during a period of time when $SO_2$ abundance exceeded $O_2$ abundance, perhaps due to an updraft (Toon et al. 1982), and that the "steady state" size distribution in the middle cloud is more akin to the reduced sulfur cases, with a low mode 1 abundance, or (3) our model is missing several pieces of essential physics, such as the microphysics of the sulfur particles, which would otherwise result in greater mode 1 abundance even when sulfur production is reduced. Option (3) is unfortunate but likely, given recent modeling results by Carlson (2010, revisions via personal communication, Sept $2^{nd}$ 2013), which showed that the ratio between total sulfur mass and total sulfuric acid mass in the cloud droplets is on the order of 0.1-1%. We can calculate the same ratio for our model by adding up the masses of the bare condensation nuclei and the sulfur cores of the sulfuric acid droplets and dividing it by the mass of the sulfuric acid in the droplets. Figure 16 shows this ratio for our three sulfur production rate cases. The sulfur-to-sulfuric-acid mass ratio steadily decreases with altitude until the middle cloud, where it quickly decreases due to the turbulence in the region causing increased condensation of sulfuric acid onto the droplets; the ratio increases below the middle cloud as sulfuric acid begins evaporating from the droplets. The best fit case is evidently one where the sulfur production rate is one order of magnitude less than that of our nominal case, but this contradicts figure 15, where that case did not yield enough mode 1 particles in the middle cloud to match the observations. This discrepancy may be avoided if sulfur microphysics were included in the model, as the small sulfur particles could then grow to larger sizes via condensation of sulfur vapor, perhaps to sizes comparable to that of mode 1 particles. In other words, the mass would be concentrated in mode 1 particles rather than the “sea” of smaller condensation nuclei, even for the cases with reduced sulfur production.

Finally, it should be noted that the periodic behavior of the nominal case disappears in the reduced sulfur cases. This is consistent with our hypothesis that the oscillation is caused by slow growth of mode 1 particles due to coagulation with the background “sea” of much smaller particles – if the “sea” is reduced, then coagulation may not be enough to allow mode 1 particles to grow past the Kelvin barrier. This further suggests that the reduced sulfur cases match the true state of the Venus clouds more closely than the nominal case, as no periodic behavior similar to those in the nominal case has yet been observed.

*3.4 Transient Wind Results*

Figure 17 gives the number density results before (black), immediately after (blue), and about an Earth week after a transient updraft event lasting ~14 Earth hours ($5 \times 10^4$ s) (red), using the wind speed profile given in Figure 6. We see that a detached haze layer forms at 75 km with a peak number density 2 orders of magnitude greater than the number density at that altitude before the wind event. The altitude of the detached haze is likely artificial given our wind profile, and the actual maximum detached haze number density is likely overestimated due to the lack of horizontal diffusion aiding (the already present) vertical diffusion in smoothing out the number density profile. However, the occurrence of an increase in number density at the altitude of the turn-over should be profile-independent. In the week that follows, the detached haze layer diffuses away so that the peak number density is more than an order of magnitude lower than its maximum immediately following the wind event. This shows that such a wind event produces the right time scales for the observed haze variability from Wilquet et al. (2012), on the order of days. Furthermore, the diffusion of the detached haze particles upwards results in an increase in the number density of the UH such that there is now a much better agreement between the Wilquet et al. (2009) UH number density observations and that of our model above 80 km. It is clear, however, that the agreement between data and model would be even better if the relaxed distribution was shifted up by 10 km. This suggests that the turnover was actually 10 km higher in the actual observations, though it is unlikely that such large gusts existed at such high altitudes.

Figure 18 shows the size distributions at altitudes close to the detached haze layer at the same times as Figure 17. Multiple size modes form below the turnover of the wind immediately following the wind event (blue). The small mode at ~0.2 μm is largely unchanged from the equilibrium distribution for altitudes below the detached haze, but is much more abundant within the detached haze itself at 75 km; this is caused by the advection of these small particles up to the altitude of the detached haze. The “bump” associated with the high altitude tail of the mode 2 cloud particles in the equilibrium distribution (black) becomes much more distinct due to upward advection of more mode 2 particles; this feature is much more prominent below the detached haze, which matches the observations of Wilquet et al. (2009) where the large UH particle mode (mode 2 cloud particles) number density drops off with altitude much faster than the small UH particle mode (mode 1 cloud particles). The small mode radii covers sizes < 0.6 μm, which is a larger range than the Wilquet et al. (2009) result of 0.1-0.3 μm; the large mode radii covers the sizes between 0.6 and 1 μm, similar to their results. Beyond 1 μm, the particle abundance drops by over an order of magnitude, though a third mode does appear resulting from advected mode 3 cloud particles; however, it is insignificant compared to the small and large modes. Above the turnover, the size distribution remains largely mono-modal.

Below the detached haze, the large mode becomes less distinct following the relaxation period (red), though both modes increase in number density. This is due to the sedimentation of large mode particles from the detached haze, as we see a large decrease in number density at the altitude of the detached haze itself. There is also an increase in number density above the turnover, as the detached haze particles diffuse upwards from the number density maximum. Thus, if the bimodality observed by Wilquet et al. (2009) was caused by winds, then it was likely transient in nature. This is supported by the order of magnitude variability between extinction profiles of the UH taken only days apart presented in Wilquet et al. (2012). It is unlikely that condensational and coagulational growth play any important role here, as their timescales are far greater than the observed variability timescale (James et al. 1997; Seinfeld and Pandis 2006).

In section 2.6 we noted the unrealistic long gust duration we have used (~14 Earth hours). However, we see that the UH size distribution is still visibly disturbed almost a full Earth week after the wind event in question. If we suppose that our wind event is caused by subsolar convection near 50° N that lasted for a full 14 hours and that the wind was constantly upwards with velocity ~1 m $s^{-1}$ at the cloud top, then given poleward meridional velocities of ~5 m $s^{-1}$ (Rossow et al. 1990) (a high estimate), it would take ~ 1 Earth week for this parcel of air to reach 70° N, the latitude where the measurements of Wilquet et al. (2009) were taken. It may not be surprising then that our "relaxed" UH number density profile matches the Wilquet et al. (2009)

number density data so well. However, most of these assumptions are highly optimistic and it is much more likely that our results show an exaggeration of what transient winds at the cloud tops can really do to the steady state aerosol distribution.

## 4. SUMMARY AND CONCLUSIONS

In this study we simulated the clouds and upper haze of Venus using version 3.0 of the microphysical and vertical transport model CARMA. We showed that appropriate choices of initial, boundary, and model atmospheric conditions can not only satisfactorily reproduce the number density and size distributions of the Venus clouds as seen in Pioneer Venus data (Knollenberg and Hunten 1980), including the bimodal and possible trimodal particle size spectrum and the three separate cloud layers, but also generate a quasi-periodically varying system rather than a system with a stable equilibrium distribution. The disagreements between our results and the observations – the overestimation of the particle number density and abundance of mode 2 particles in the middle cloud and the underestimation of the lower cloud and the abundance of mode 3 particles – are all within the range of variability seen in other observations, though the addition of transient gusts in the middle cloud, which would both deplete the middle cloud of particles and create a greater number of larger particles, may improve our agreement with the Pioneer Venus Sounder Probe data.

We also simulated the upper haze as a mixture of droplets formed from *in situ* nucleation of sulfuric acid vapor on meteoric dust and droplets upwelled from the cloud decks below. We showed that the former population rapidly coagulates with the latter population, resulting in a mono-modal size distribution. Meanwhile, for altitudes below 80 km there is good agreement between our model particle number densities and the sum of the small mode (mode 1) and large mode (mode 2) number densities from Wilquet et al. (2009); above 80 km, we underestimate the number density by about half an order of magnitude. These particles are likely those originally observed by the Pioneer Venus OCPP (Kawabata et al. 1980), and could represent the haze under stable atmospheric conditions. These discrepancies were reduced upon the application of a transient updraft, which created a detached haze layer at the altitude of turnover. The diffusion of the detached haze upwards increased the particle number density of the UH such that the results agreed with the UH number densities derived by Wilquet et al. (2009) above 80 km. The resulting size distribution showed a clear bimodal structure below the detached haze immediately after the wind event, with the small mode (mode 1) particles having radii < 0.6 μm, and the large mode having radii between 0.6 and 1 μm, similar to the results of Wilquet et al. (2009). The relaxation of the multi-modal structures in the size distribution back to a mono-modal distribution had a timescale of roughly an Earth week, similar to the time scales of haze variability observed by Venus Express (Luz et al. 2011, Markiewicz et al. 2007). The duration and strength of the updraft were likely exaggerations of what actually takes place in the Venus atmosphere, though subsolar convection could be a viable candidate for the generation of these events under optimistic assumptions.

We noted a quasi-periodic variability in our results with a rough period of ~6 Earth months despite a stable background atmospheric state and constant input of sulfuric acid vapor and condensation nuclei. This variability is most apparent in the middle cloud, where steady coagulation of small particles (~10 nm) with mode 1 particles lead to a slow increase in the mean mode 1 particle radii. Upon reaching the Kelvin barrier, the mode 1 particles rapidly grow to mode 2 particles via condensational growth, while upward diffusion of mode 1 particles from below replenish the mode 1 population. The cycle then repeats and a new wave of mode 2 particles appear; these new mode 2 particles coagulate with the old mode 2 particles, resulting in mode 3 particles that then sediment out of the model domain. This process does not occur if the sulfur condensation nuclei production rate is decreased by an order of magnitude or more, nor does it occur if the condensation nuclei are larger. Comparison between model results and Pioneer Venus observations (Knollenberg and Hunten 1980) suggests that the LCPS data may have been taken during one of the unstable "growth phases" rather than the much more long-lived stable phases. These episodic variations also leads to periodicity in the magnitude of the sulfuric acid flux out of the bottom boundary of the model domain and suggests possible sulfuric acid "rain events" that could occur immediately below the cloud deck. Such long term variability in particle population should be detectable by VIRTIS observations of the Venus night side.

The sulfur production rate in the upper cloud is uncertain and could be much lower than half of the sulfuric acid production rate as assumed in Imamura and Hashimoto (2001) and in our nominal model. Comparisons with sulfur-to-sulfuric-acid mass ratios of the cloud droplets derived from UV data (Carlson 2010, revisions via personal communications, Sept 2nd 2013) shows that the best fit case is one where the sulfur production rate is reduced by an order of magnitude from that of the nominal case. However, comparison of this case with LCPS particle size distributions shows that the model predicts mode 1 particle abundances in the middle cloud two orders of magnitude less what is observed. This discrepancy could be caused by the lack of sulfur microphysics in our model, which could allow the sulfur condensation nuclei to grow to mode 1 particles (~0.2 μm) instead of being fixed at a size of ~10 nm.

It is necessary for future models to include both sulfur and sulfuric acid microphysics in order to provide a more complete picture of the processes occurring within the Venus clouds and hazes. Interactions between aerosols formed from these species and their gas phases are diverse and only a subset has been tested. Furthermore, the production rates of these two species, particularly sulfur and its allotropes, require better constraints from both chemical models that take into account sinks due to aerosol formation and observations that are able to probe down to the altitudes where photochemistry dominates. Finally, continuous observation of the Venus hazes and clouds is essential in constraining their time evolution, especially after transient events.

## Acknowledgements

We thank S. Garimella and R. L. Shia for assistance with the setting up and running of the CARMA code. We thank R. W. Carlson and C. Parkinson for their valuable inputs. We thank C. Li for his help in speeding up our model runs by more than a factor of 10. This research was supported in part by the Venus Express program via NASA NNX10AP80G grant to the California Institute of Technology, and in part by an NAI Virtual Planetary Laboratory grant from the University of Washington to the Jet Propulsion Laboratory and California Institute of Technology. Part of the research described here was carried out at the Jet Propulsion Laboratory, California Institute of Technology, under a contract with the National Aeronautics and Space Administration.

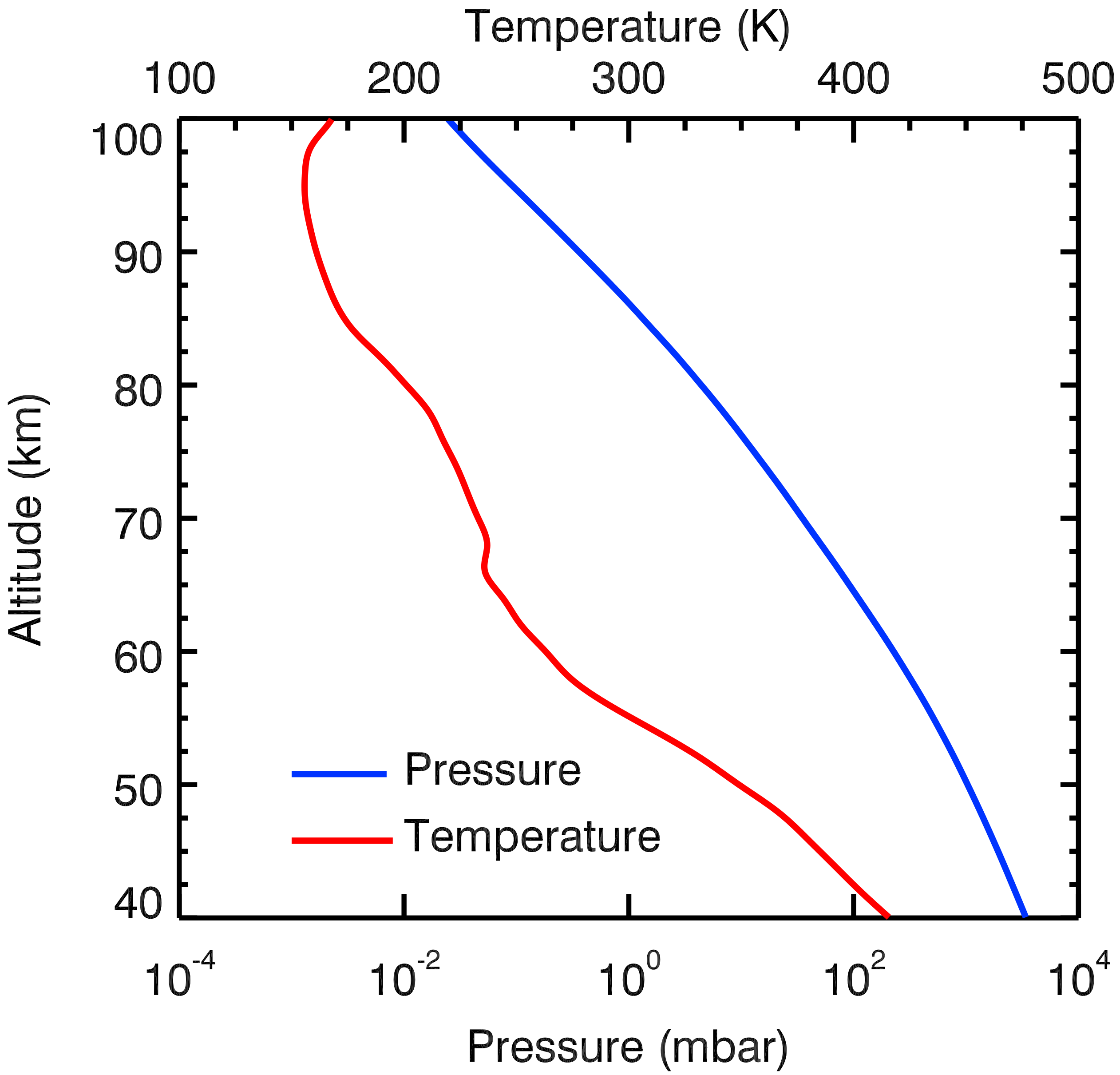

Figure 1. Model temperature (red) and pressure (blue) profiles taken from the Venus International Reference Atmosphere (Seiff et al. 1985).

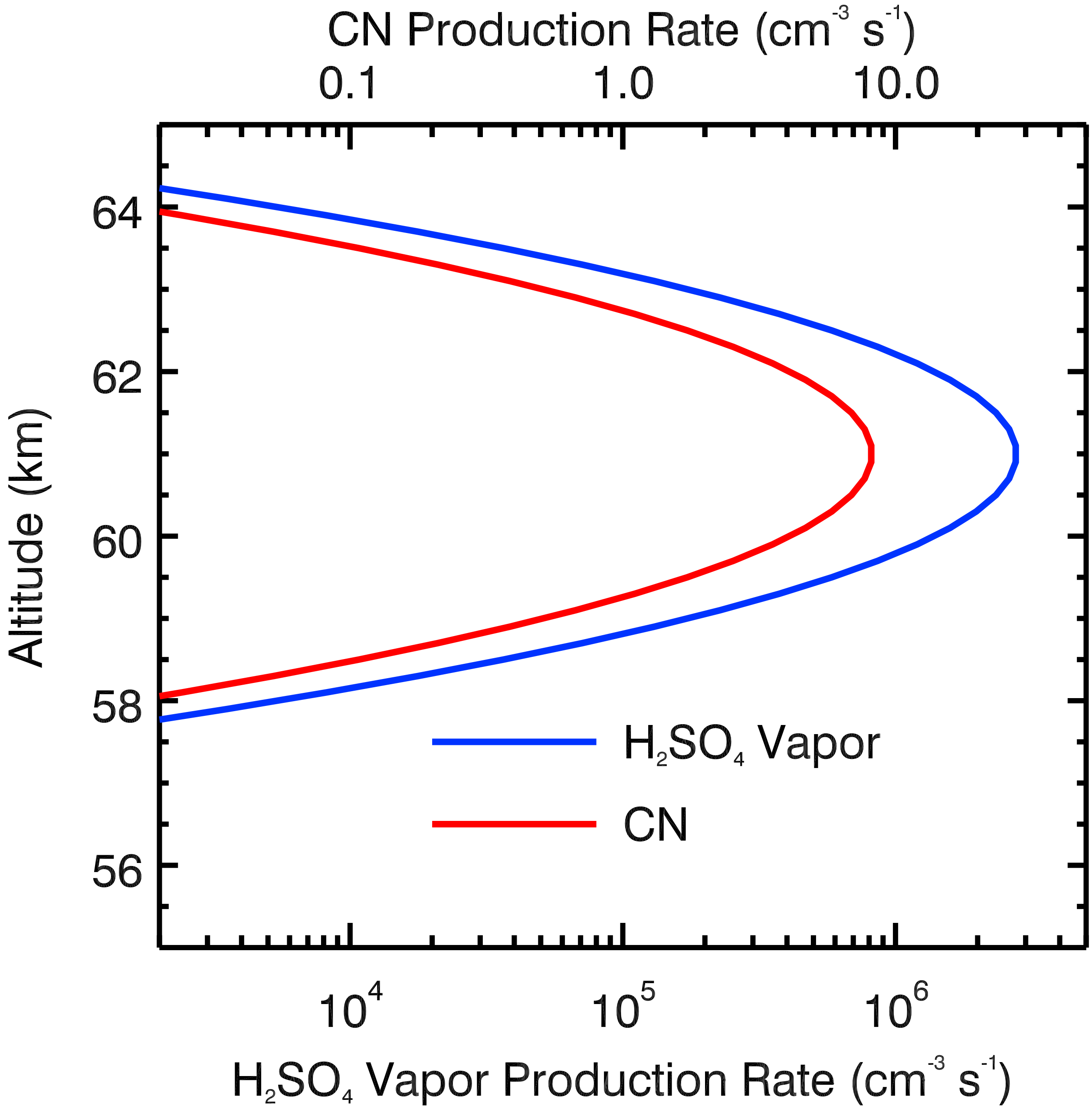

Figure 2. Model production rate profiles for sulfuric acid vapor (blue) and photochemical condensation nuclei (red), based on that of Imamura and Hashimoto (2001) with the peak rates adjusted to fit LCPS data.

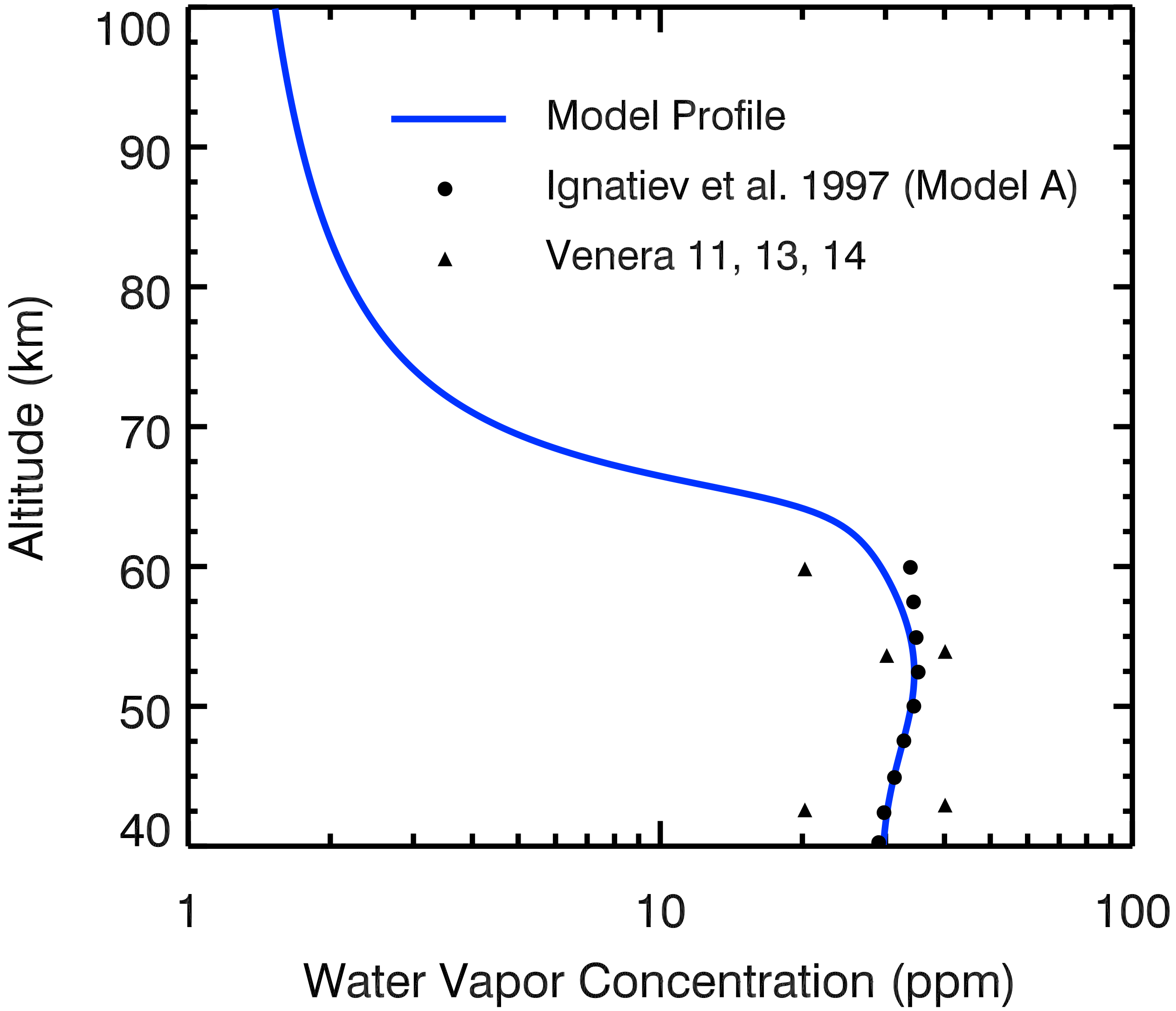

Figure 3. Model water vapor profile (blue) plotted with the Model A (filled circles) and Venera 11, 13, and 14 data (triangles) from Ignatiev et al. (1997). The water vapor concentration in the upper haze is taken to be ~1 ppm from observations by Bertaux et al. (2007).

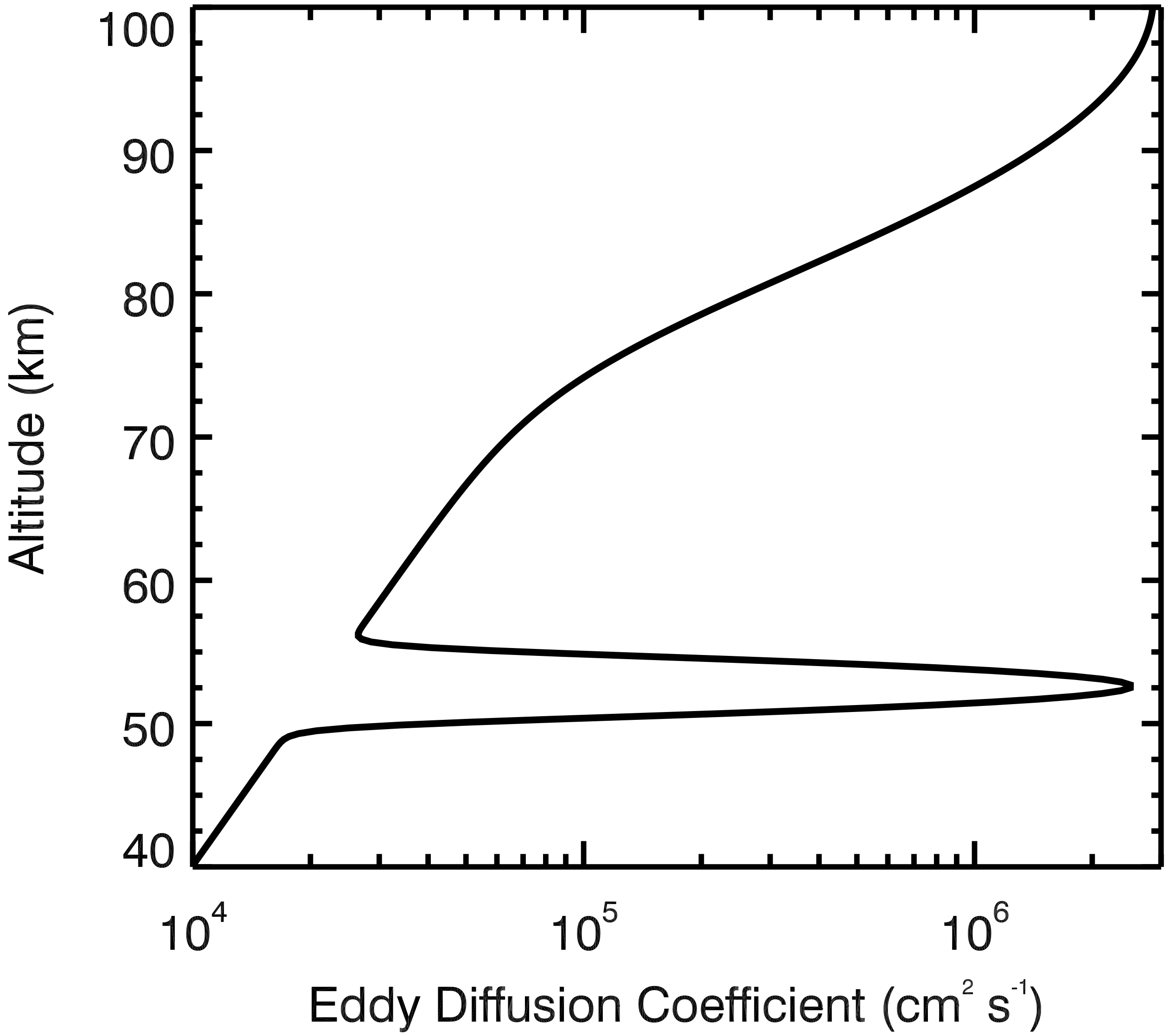

Figure 4. Model eddy diffusion coefficient profile, with the 40-70 km section based on Imamura and Hashimoto (2001), and the 70-100 km section based on Krasnopolsky (1983).

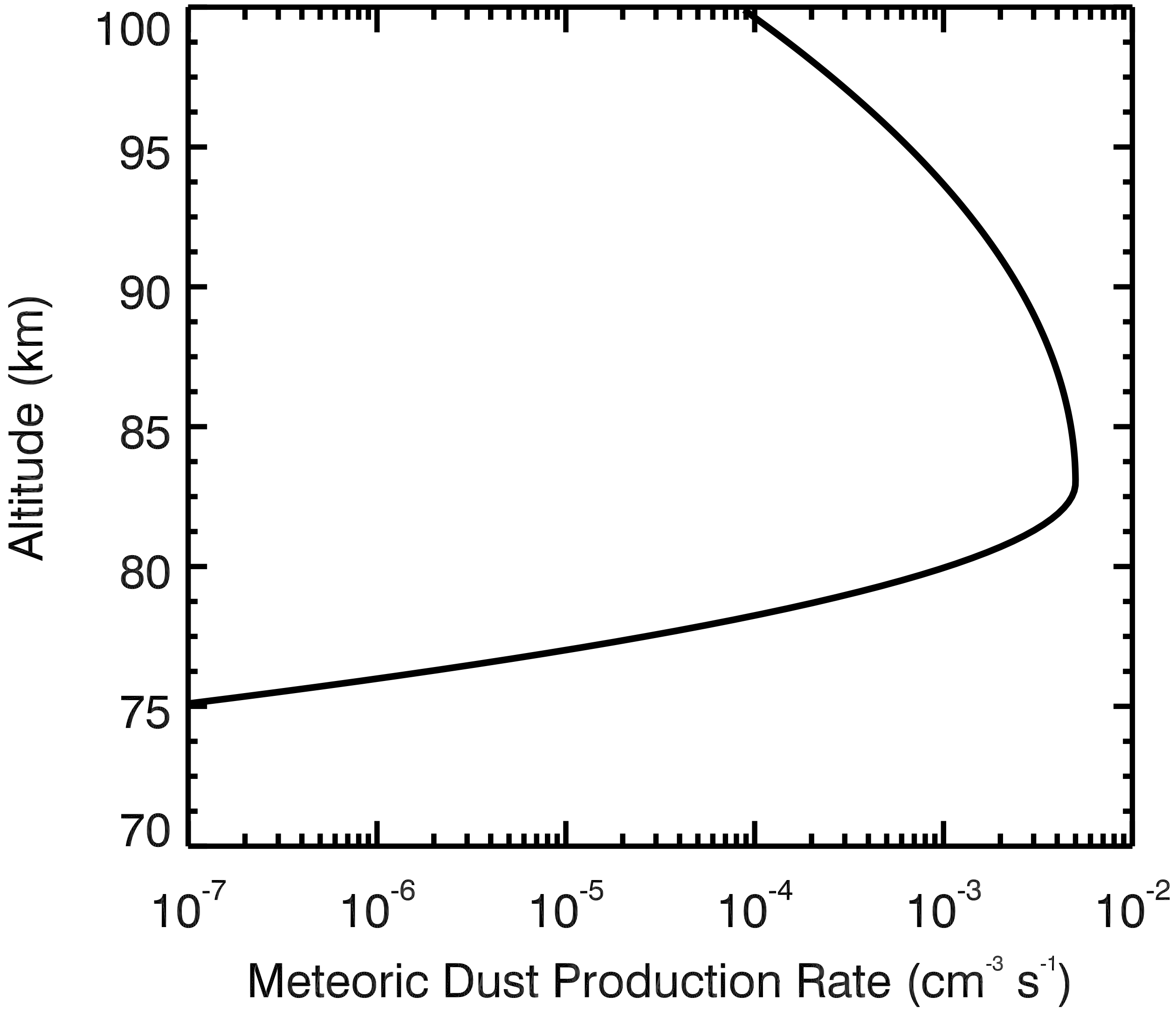

Figure 5. Model meteoric dust production rate profile, based on Kalashnikova et al. (2000), normalized to 1.3 nm particles, and shifted down from the original distribution by 4 km in order for the maximum of this profile to match that of the number density profile of the small mode particles in the UH, as retrieved from solar occultation data by Wilquet et al. (2009).

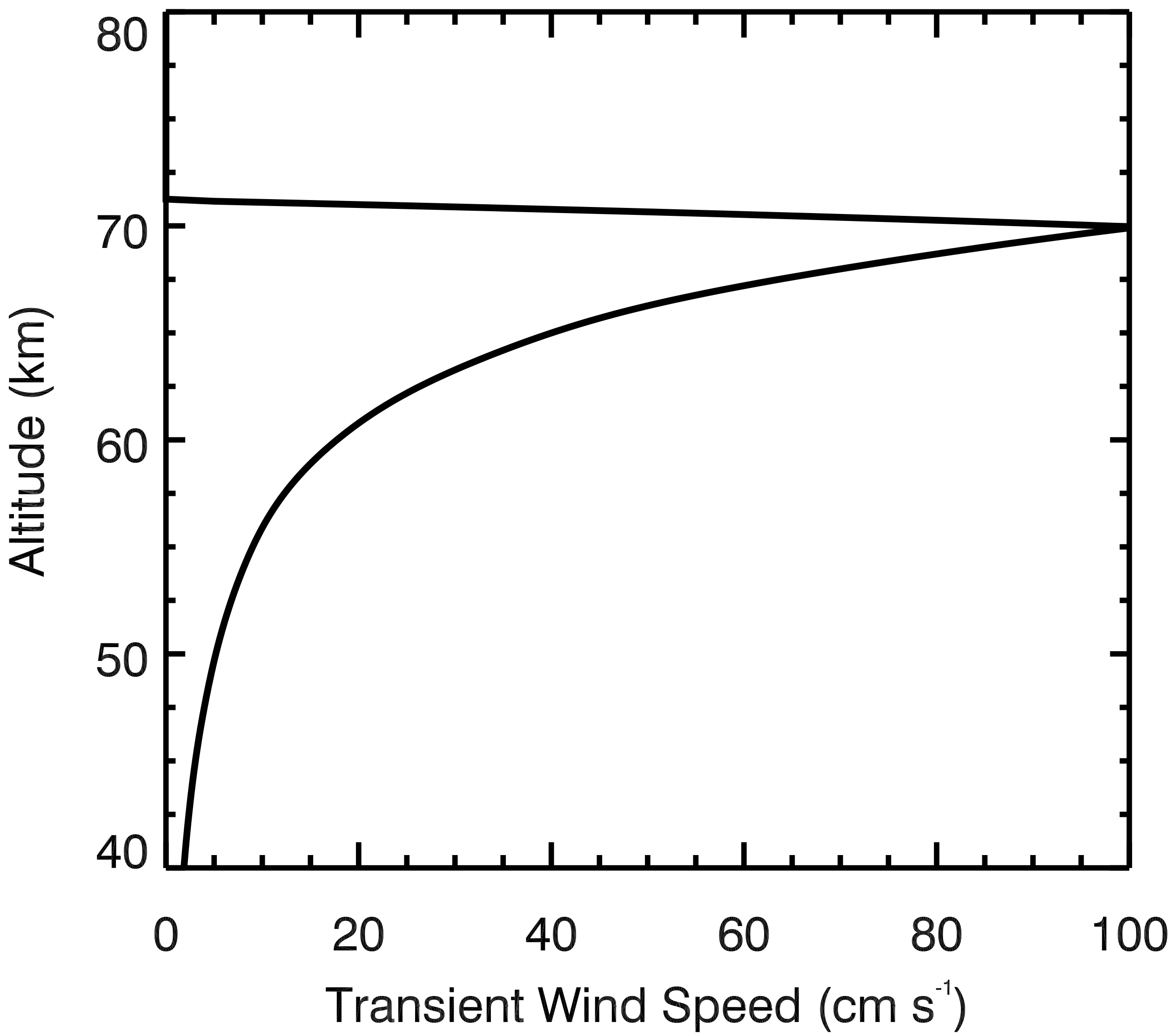

Figure 6. Model wind speed profile with the portion below 70 km taken from Imamura and Hashimoto (2001), and the cut-off above 70 km representing the turning over of the upwelling.

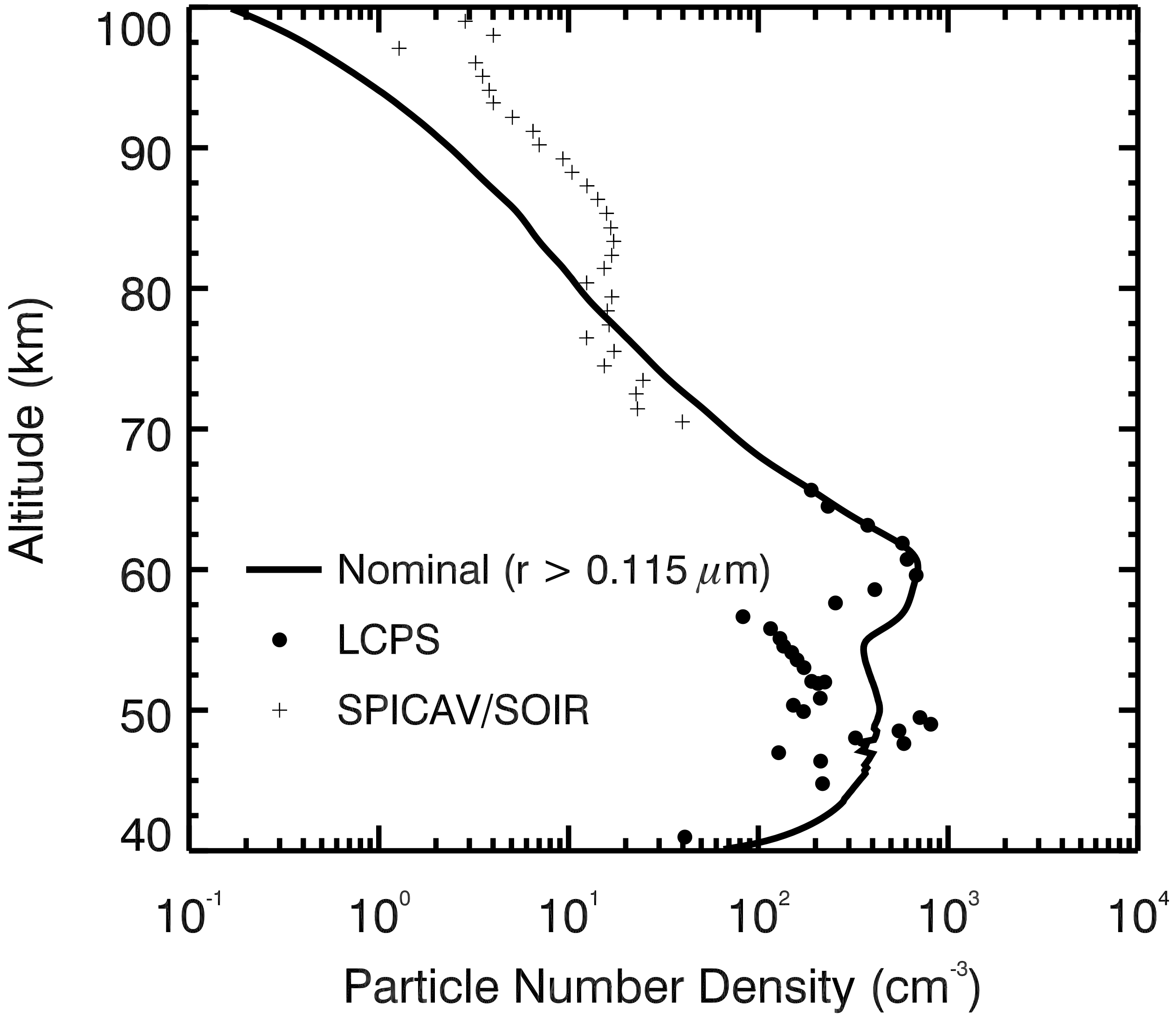

Figure 7. Number density of cloud and haze particles with radius r > 0.115 μm (solid line) from the nominal model compared to total number density data from LCPS (filled circles) (Knollenberg and Hunten 1980) and Venus Express (pluses) (Wilquet et al. 2009).

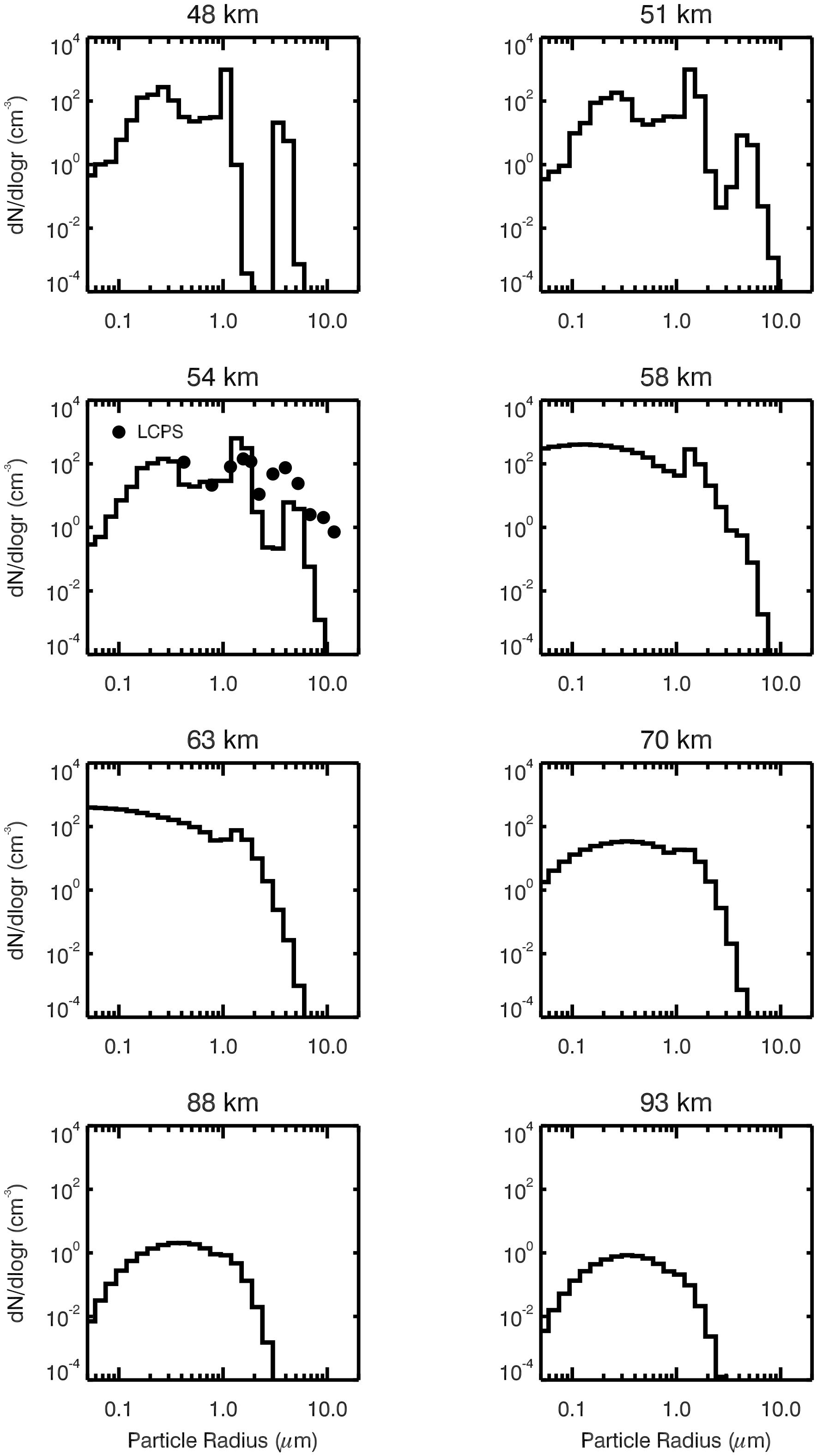

Figure 8. Particle size distributions at various altitudes from the nominal model. LCPS size data at 54.2 km (dots) (Knollenberg and Hunten 1980) is plotted for comparison.

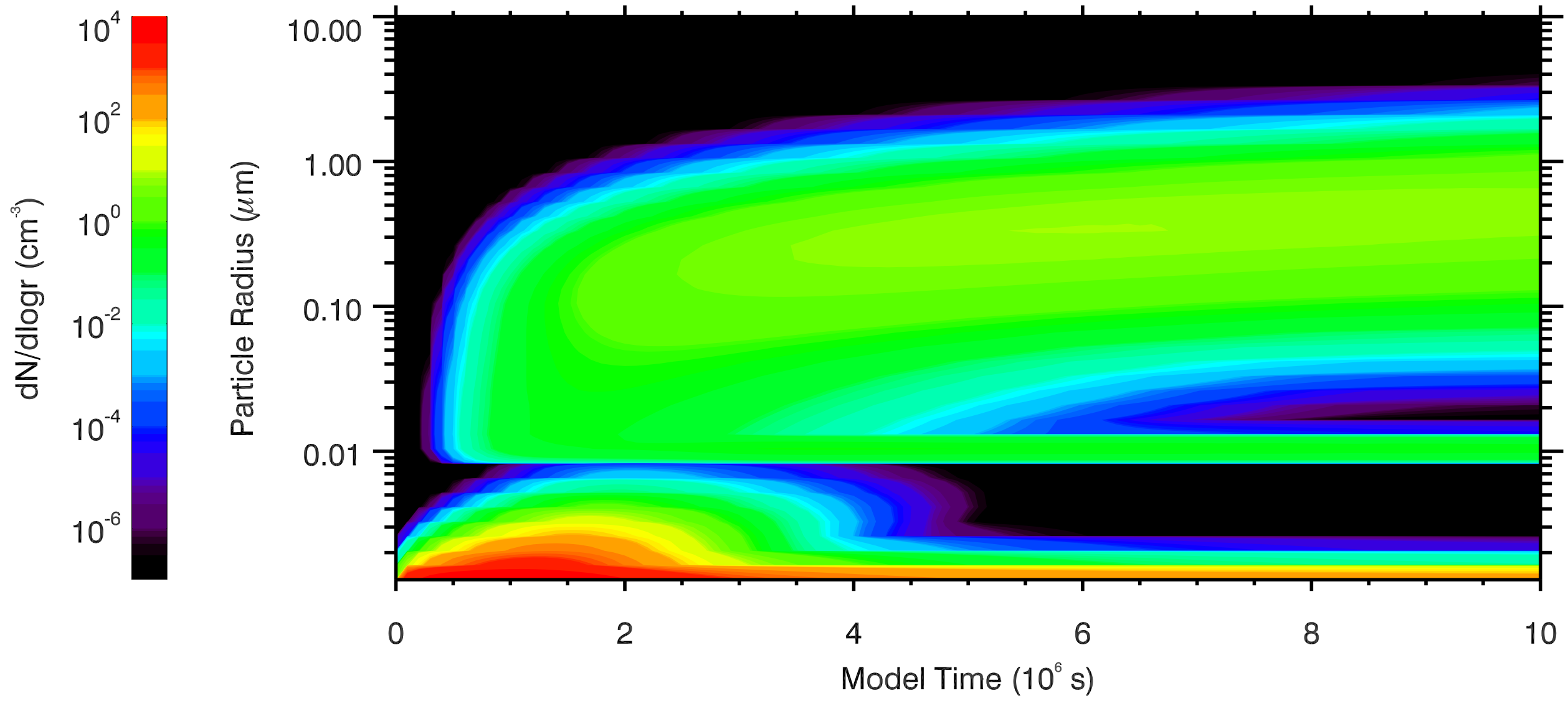

Figure 9. Time evolution of the UH particle size distribution at 84 km at the beginning of the nominal model run, from t = 0 to t = $10^7$ s. The green "bar" at 10 nm is a result of the artificial injection of 10 nm particles into the model domain as photochemical condensation nuclei and should be ignored. The black regions represent parts of the phase space where dN/dlogr < $10^{-7}$ $cm^{-3}$.

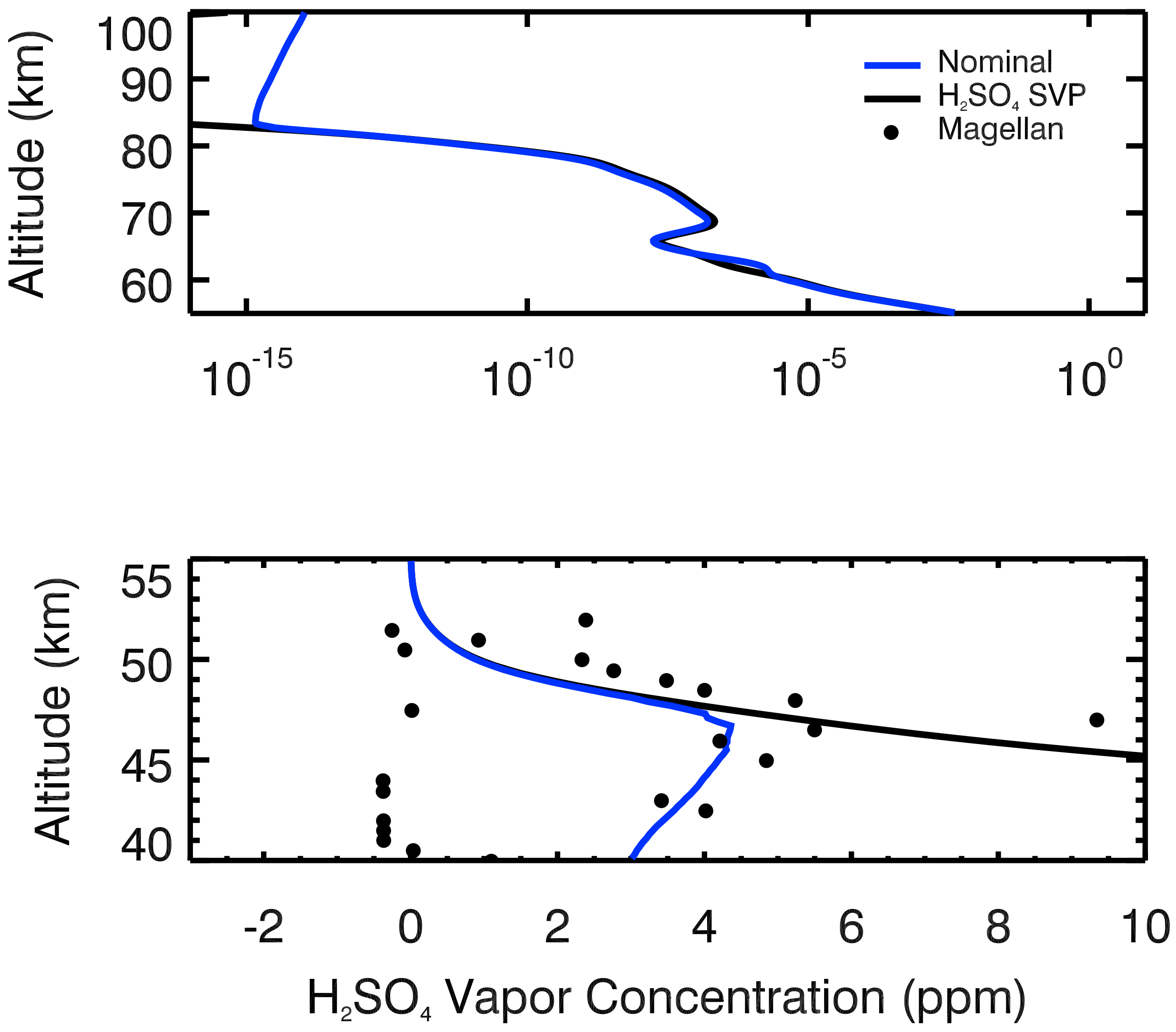

Figure 10. Sulfuric acid vapor mixing ratio from the nominal model (blue) compared with the sulfuric acid saturation vapor pressure over a flat surface (black) (sec. 2.2) and Magellan radio occultation data analyzed by Kolodner and Steffes (filled circles) (1998).

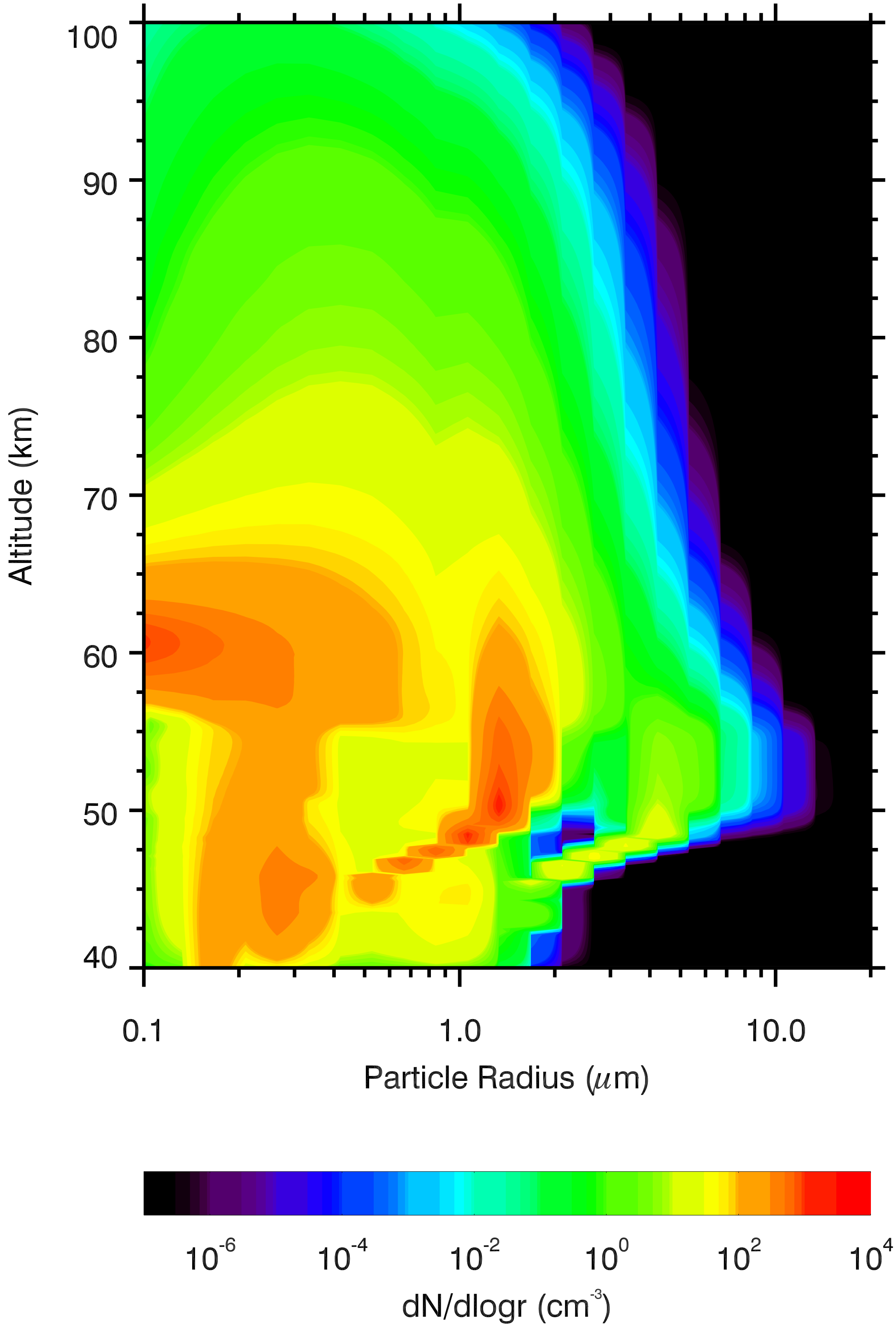

Figure 11. The nominal particle number density as a function of particle size and altitude. The black regions represent parts of the phase space where dN/dlogr < $10^{-7}$ $cm^{-3}$.

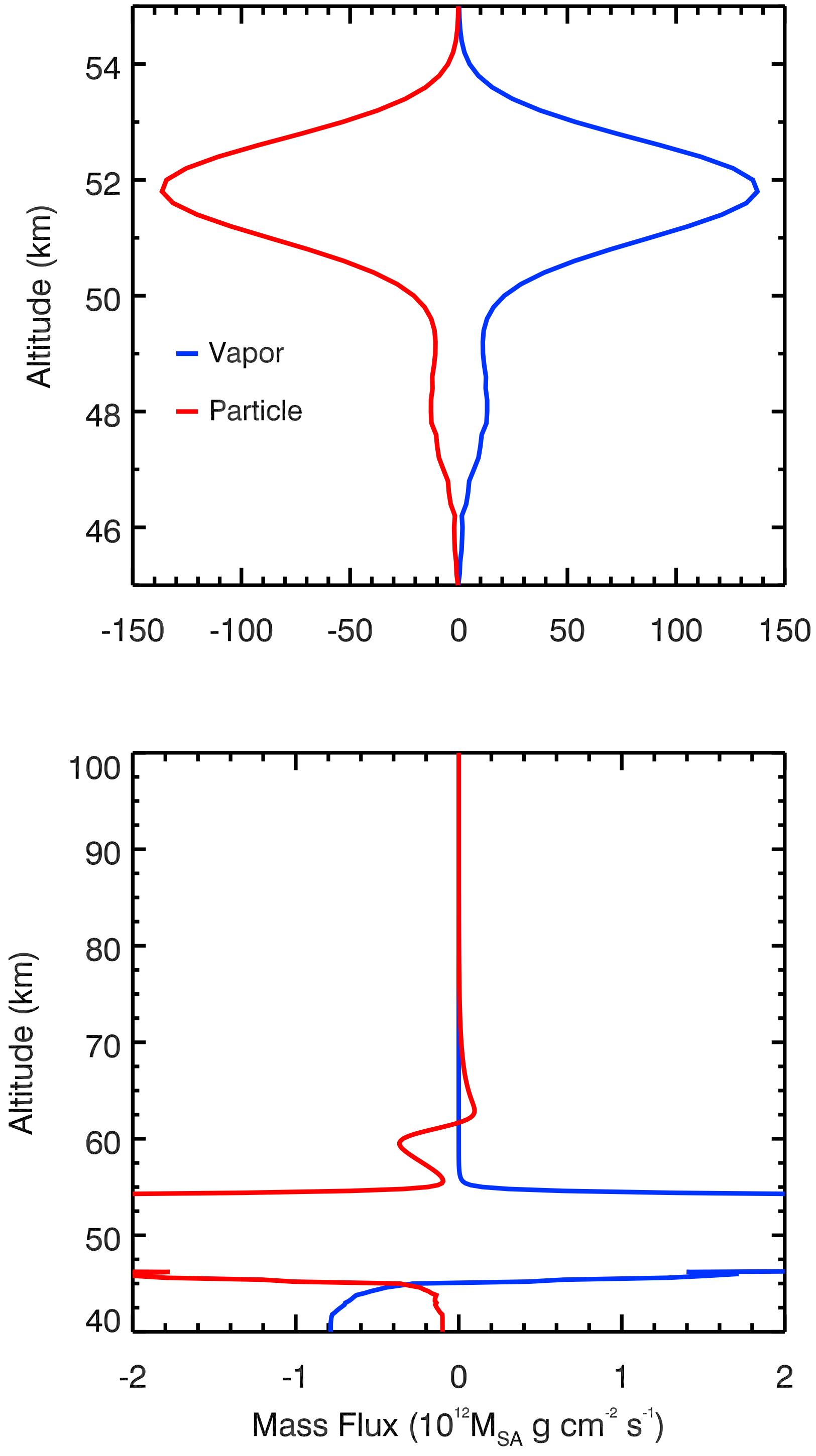

Figure 12. The nominal sulfuric acid vapor (blue) and particle (red) mass fluxes at steady state at the same time step as in figures 7, 8, 10, and 11 expressed in units of mass equivalent to $10^{12}$ sulfuric acid molecules per unit area per second, where each molecule has mass $M_{SA} \sim 1.6 \times 10^{-22}$ g. Note the different axes scales between the top and bottom panels: the top panel shows the high flux values of the middle cloud, while the bottom panel shows the lower flux values at the other altitudes.

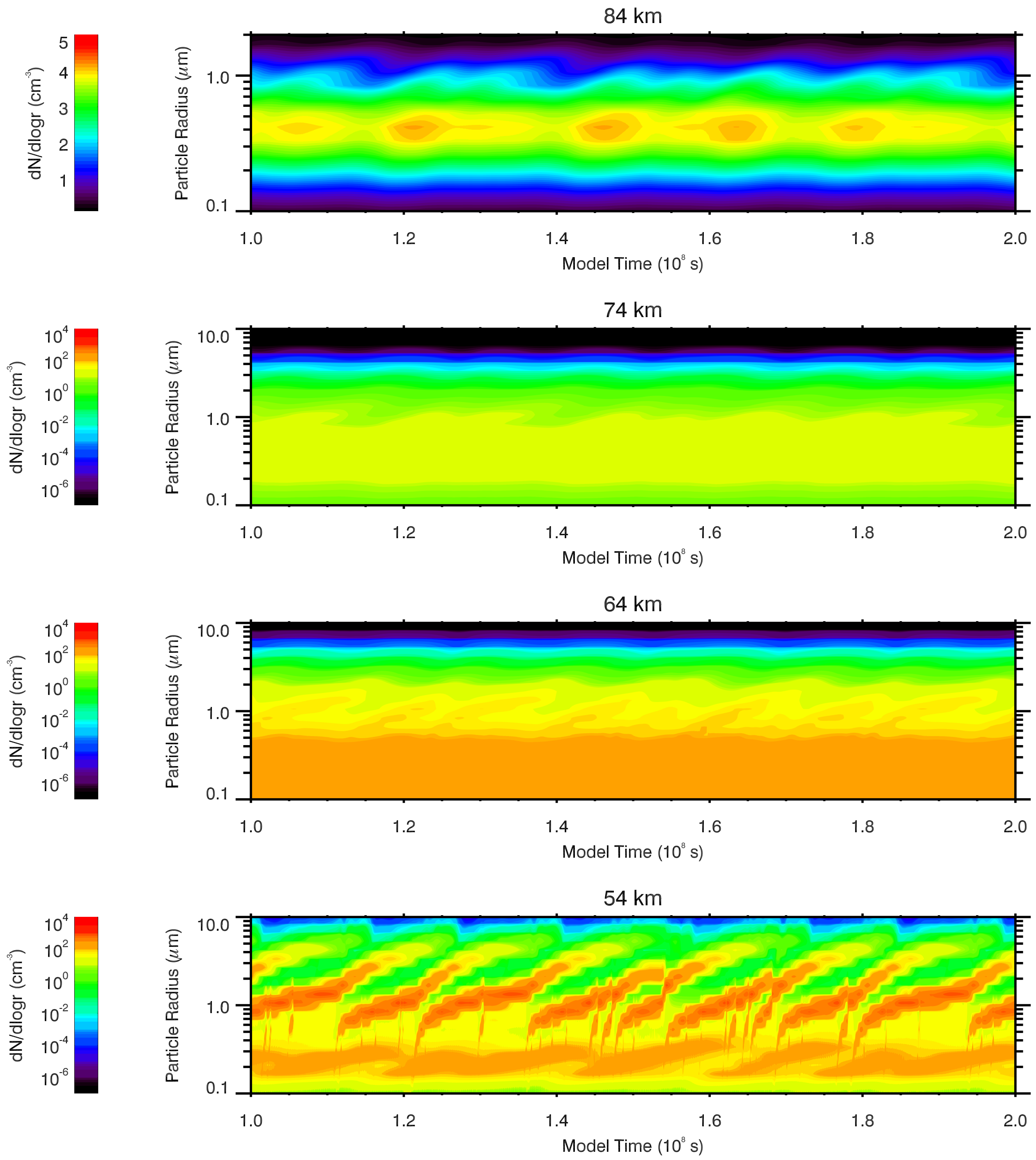

Figure 13. Time evolution of the nominal particle size distribution at 84, 74, 64, and 54 km from $t = 10^8$ s to $t = 2 \times 10^8$ s. Note the different number density contour and y axis scale for the 84 km plot. The black regions represent parts of the phase space where dN/dlogr < $10^{-7}$ $cm^{-3}$ at 74, 64, and 54 km, and < 0.1 $cm^{-3}$ at 84 km.

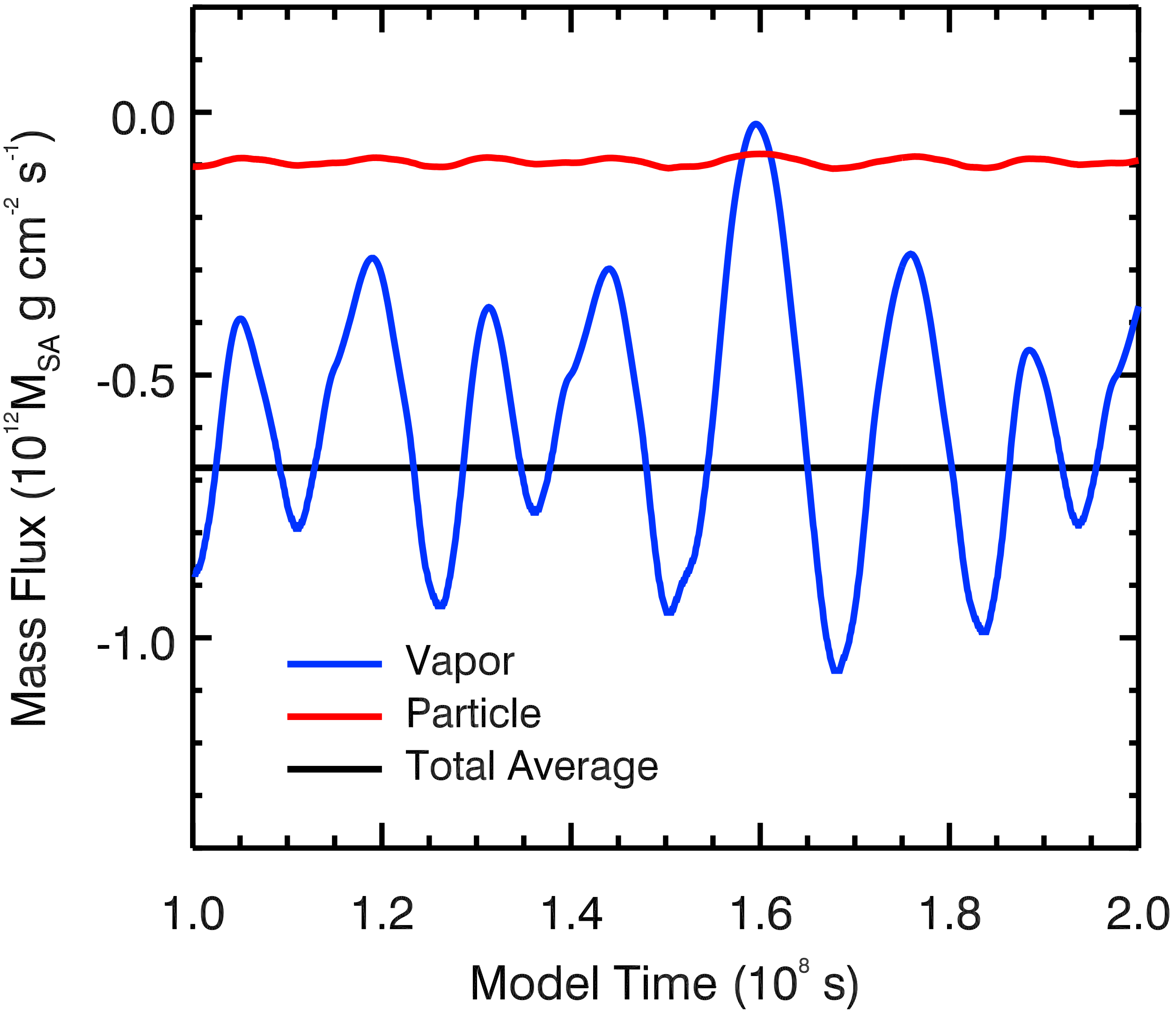

Figure 14. The time evolution of the nominal sulfuric acid vapor (blue) and particle (red) mass fluxes at the bottom of the model domain from t = $10^8$ s to t = 2 x $10^8$ s plotted with the average of the total flux during this time period (black), all expressed in units of mass equivalent to $10^{12}$ sulfuric acid molecules per unit area per second, where each molecule has mass $M_{SA} \sim 1.6 \times 10^{-22}$ g. The negative values indicate downward fluxes.

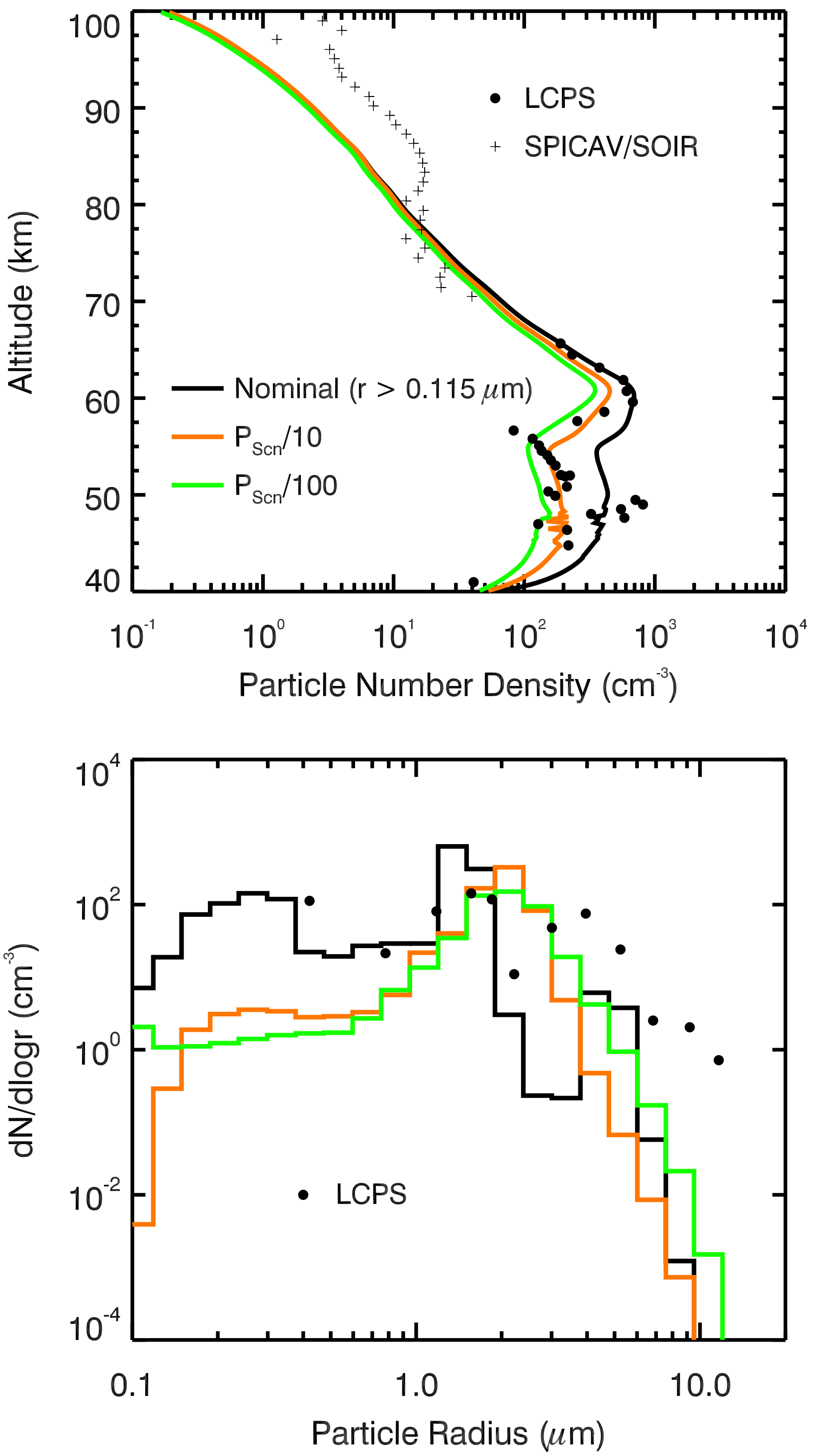

Figure 15. The number density (top) and size distribution at 54 km (bottom) of the nominal (black), one order of magnitude reduction in sulfur production (orange), and two orders of magnitude reduction in sulfur production (green) cases. The curves in the top figure are compared to total number density data from LCPS (filled circles) (Knollenberg and Hunten 1980) and Venus Express (pluses) (Wilquet et al. 2009). The histograms in the bottom figure are compared to LCPS size data at 54.2 km (filled circles) (Knollenberg and Hunten 1980).

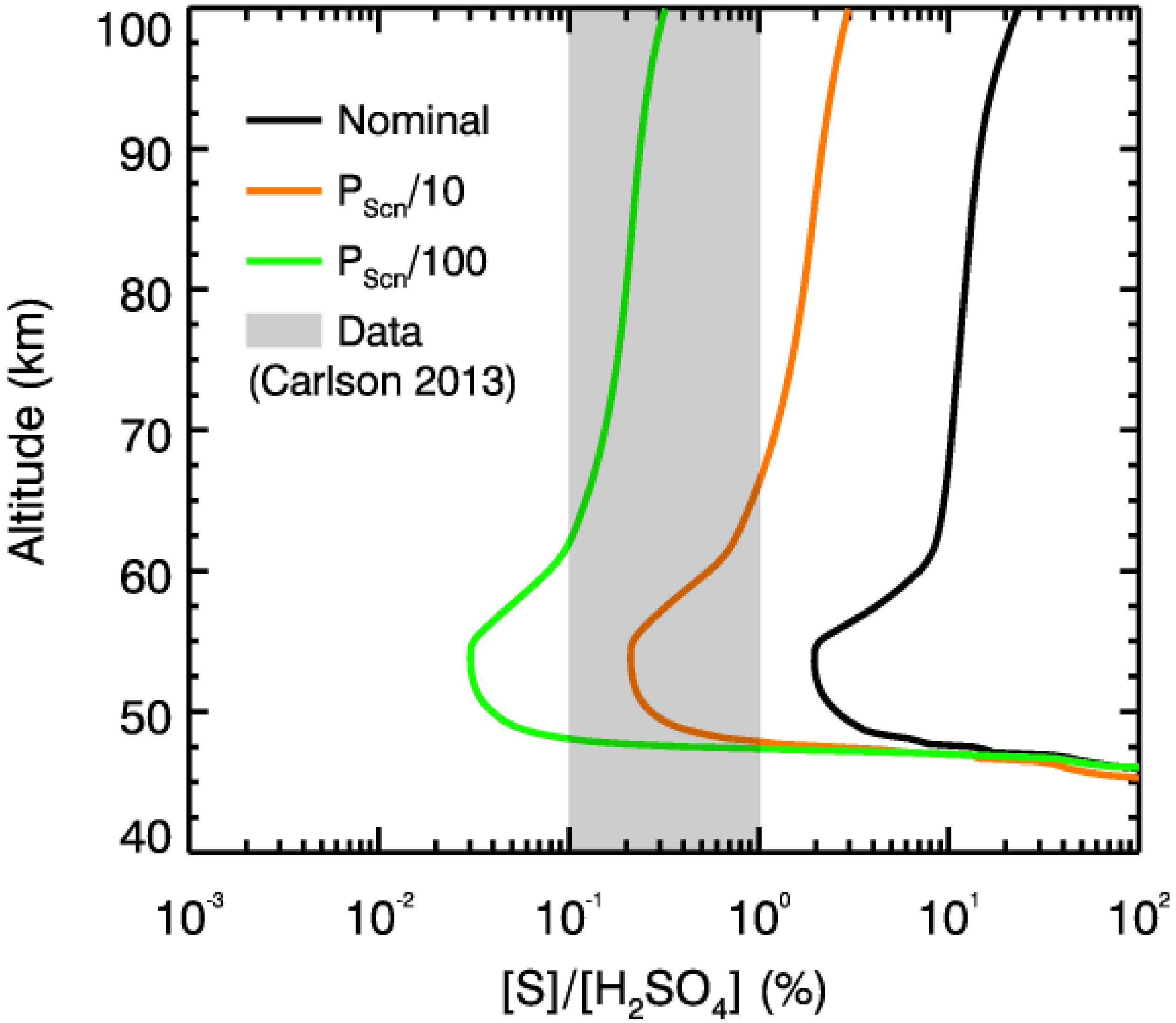

Figure 16. The average mass ratio of sulfur to sulfuric acid in cloud and haze droplets as a function of altitude for the nominal (black), one order of magnitude reduction in sulfur production (orange), and two orders of magnitude reduction in sulfur production (green) cases. The two red dashed lines show the range of mass ratios as constrained by fits to UV data (Barker et al. 1975) produced by Carlson (2010, revisions via personal communication, Sept. 2nd, 2013).

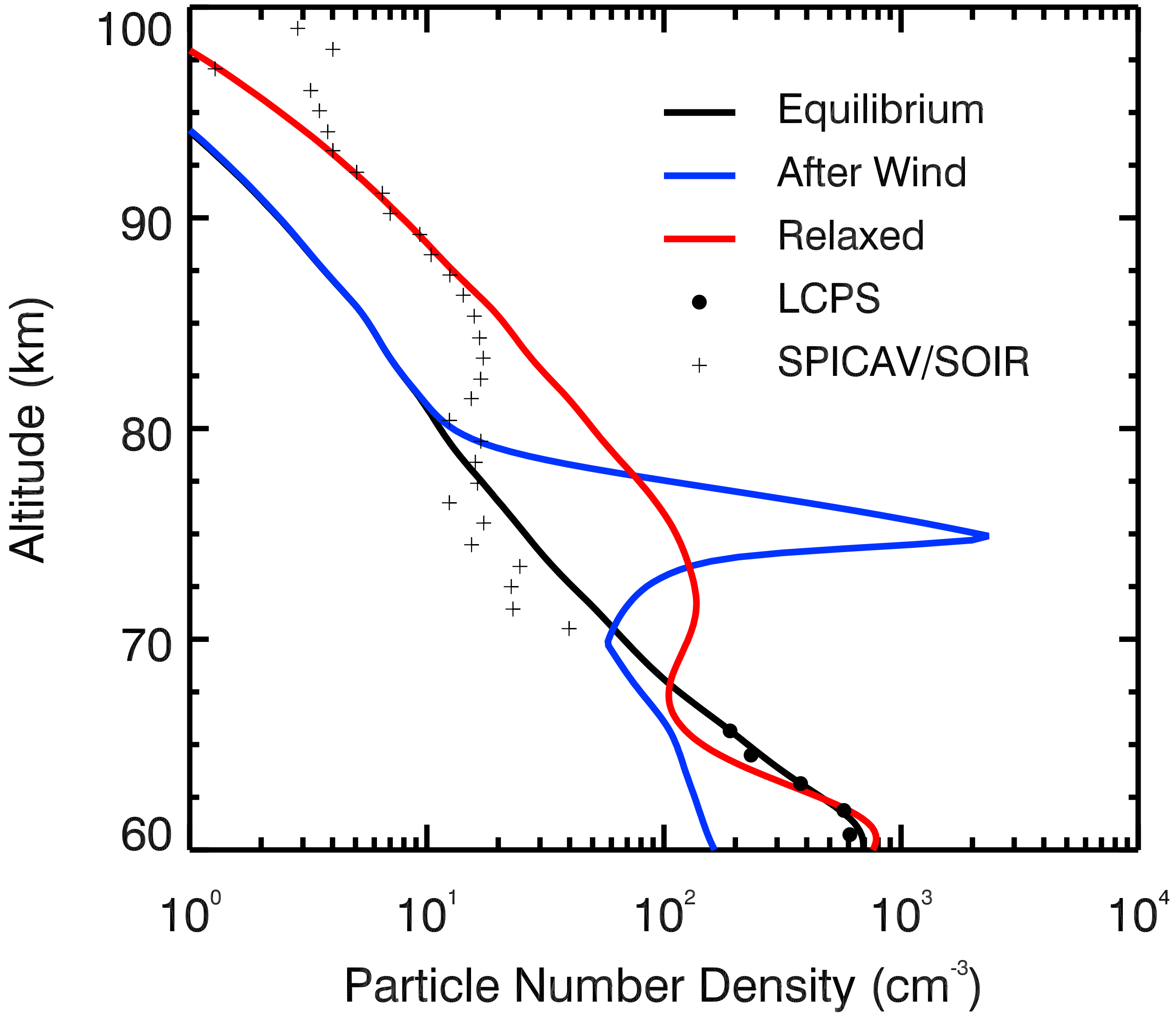

Figure 17. Number density profiles of the upper cloud and haze before (black), immediately after (blue), and 5 x $10^5$ s after (red) a 5 x $10^4$ s transient wind event. The total number density data from LCPS (filled circles) (Knollenberg and Hunten et al. 1980) and Venus Express (pluses) (Wilquet et al. 2009) are plotted for comparison. The wind speed profile is shown in Fig. 6.

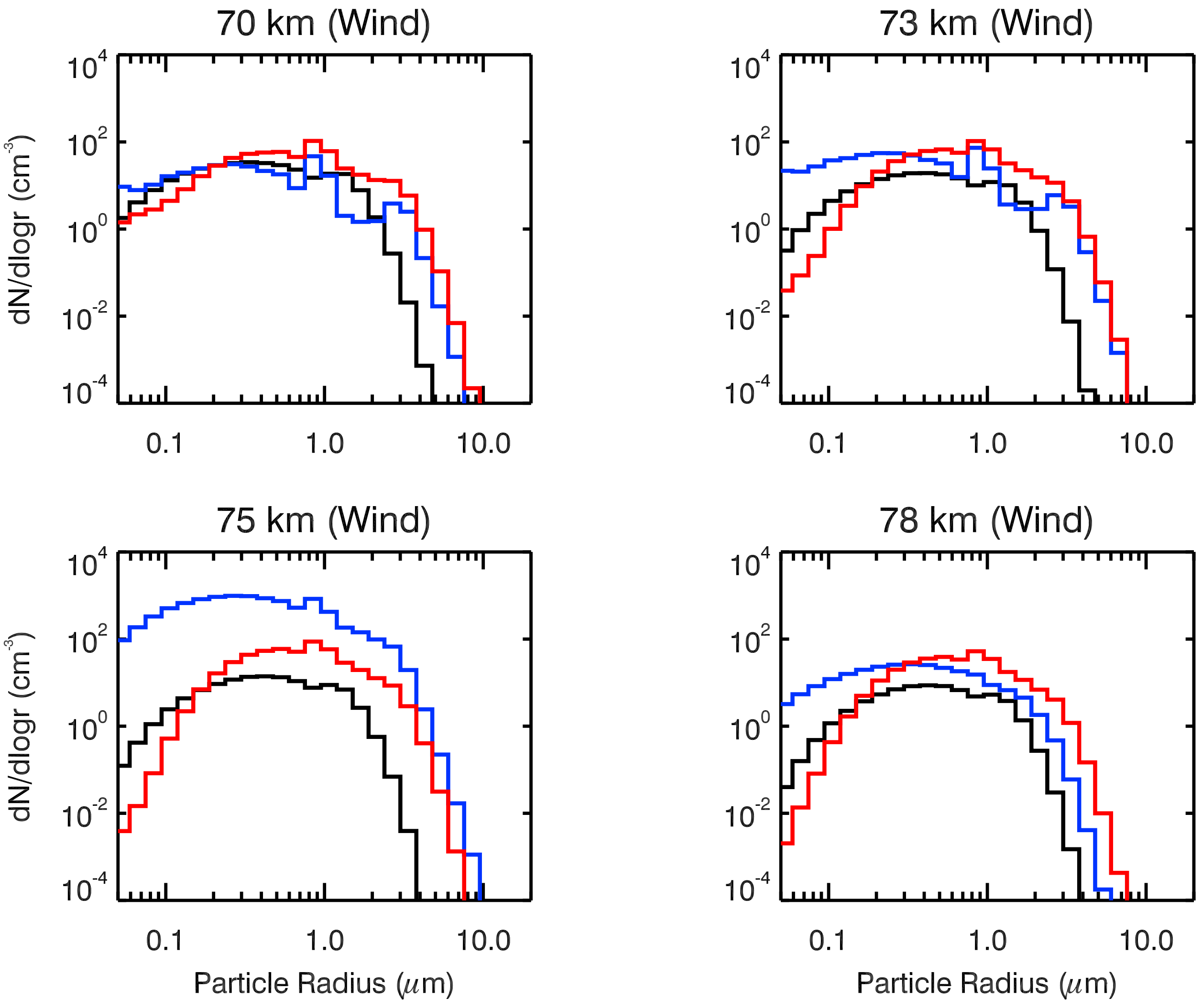

Figure 18. Particle size distribution before (black), immediately after (blue), and 5 x $10^5$ s after (red) a 5 x $10^4$ s transient wind event, plotted for various altitudes close to the turnover altitude of the transient wind. The wind speed profile is shown in Fig. 6.

Table 1. Model parameters.

| | Nominal Model | Other Values Used |
|---|---|---|
| **Surface Gravity** | 887.0 cm $s^{-2}$ | |
| **Atmospheric Mole. Wt.** | 43.45 g $mol^{-1}$ ($CO_2$) | |
| **Condensable Mole. Wt.** | 98.08 g $mol^{-1}$ ($H_2SO_4$) | |
| **Atmospheric Viscosity**[a] | 1.496 x $10^{-4}$ g $cm^{-1}$ $s^{-1}$ | |
| **Sulfuric Acid Surface Tension**[b] | 72.4 erg $cm^{-2}$ | |
| **T-P Profile** | Figure 1 | |
| **Water Vapor Profile** | Figure 3 | |
| **Production Rates** | | |
| Meteoric Dust | 4800 $cm^{-2}$ $s^{-1}$ (Fig. 5) | |
| Photochemical CNs | 1.75 x $10^{6}$ $cm^{-2}$ $s^{-1}$ (Fig. 2) | Nominal/10[c], Nominal/100[c] |
| Sulfuric Acid Vapor | 6 x $10^{11}$ $cm^{-2}$ $s^{-1}$ (Fig. 2) | |
| **Time Domain** | | |
| Time Step | 10 s | |
| Total Simulation Time | 2 x $10^{8}$ s | 5 x $10^{4}$ s[d], 5 x $10^{5}$ s[d] |
| **Altitude Domain** | | |
| Thickness of Altitude Level | 200 m | |
| Total Altitude Levels | 300 | |
| **Size Domain** | | |
| Mass Ratio Between Bins | 2 | |
| Number of Bins | 45 | |
| Smallest Bin Size | 1.3 nm | |
| Photochemical CN Size | 10.4 nm[e] | |
| **Initial Conditions** | | |
| Meteoric Dust | 0 $cm^{-3}$ | |
| Photochemical CNs | 0 $cm^{-3}$ | |
| Sulfuric Acid Vapor | 0 ppm | |
| **Boundary Conditions** | | |
| Meteoric Dust (Top) | Zero flux | |
| Photochemical CNs (Top) | Zero flux | |
| Sulfuric Acid Vapor (Top) | Zero flux | |
| Meteoric Dust (Bottom) | 0 $cm^{-3}$ | |
| Photochemical CNs (Bottom)[f] | 40 $cm^{-3}$ | |
| Sulfuric Acid Vapor (Bottom) | 3 ppm | |
| **Wind Speed** | 0 cm $s^{-1}$ | Figure 6[d] |

[a] Value at 300 K. Viscosity dependence on temperature is calculated using Sutherland's equation and parameters from White (1991).

[b] Value for a flat surface at ~60 km, where temperature ~ 260 K and sulfuric acid weight percentage of sulfuric acid droplets ~ 80%.

[c] See Sec. 3.3.

[d] For transient wind tests (Sec. 2.6 and 3.4).

[e] We use 10.4 nm instead of 10 nm as 10.4 nm corresponds to the radius represented by one of the discrete size bins in our model (bin 10).

[f] These photochemical CNs have radius ~0.17 μm, consistent with the mean size of the particles at ~40 km observed by Pioneer Venus LCPS (Knollenberg and Hunten 1980).